\documentclass[trackchanges]{aastex701}
\usepackage{amsmath}
\usepackage{graphicx}
\usepackage{subcaption} 

\newcommand{\gaia}{\textit{Gaia}}
\newcommand{\msol}{{\rm M_\odot}}

\begin{document}

\title{High-Priority Compact-Object Candidates in Young Clusters from the \textit{Gaia} DR3 Non-Single-Star Catalog}

\author[]{Dejian Liu}
\affiliation{College of Mathematics and Physics, China Three Gorges University, Yichang 443000, People’s Republic of China;}
\affiliation{Center for Astronomy and Space Sciences, China Three Gorges University, Yichang 443000, People’s Republic of China;}
\email[]{}

\author[]{Kangteng Chang}
\affiliation{College of Mathematics and Physics, China Three Gorges University, Yichang 443000, People’s Republic of China;}
\affiliation{Center for Astronomy and Space Sciences, China Three Gorges University, Yichang 443000, People’s Republic of China;}
\email[]{}

\author[]{Yao Huang}
\affiliation{College of Mathematics and Physics, China Three Gorges University, Yichang 443000, People’s Republic of China;}
\affiliation{Center for Astronomy and Space Sciences, China Three Gorges University, Yichang 443000, People’s Republic of China;}
\email[]{}

\author[]{Yingjie Li}
\affiliation{Purple Mountain Observatory, Chinese Academy of Sciences, Nanjing 210008,People’s Republic of China;}
\affiliation{University of Science and Technology of China, Jinzhai Road, Hefei 230026, People’s Republic of China;}
\email[]{}

\author[]{Xianjin Shen}
\affiliation{Purple Mountain Observatory, Chinese Academy of Sciences, Nanjing 210008,People’s Republic of China;}
\affiliation{University of Science and Technology of China, Jinzhai Road, Hefei 230026, People’s Republic of China;}
\email[]{}

\author[]{Zehao Lin}
\affiliation{Purple Mountain Observatory, Chinese Academy of Sciences, Nanjing 210008,People’s Republic of China;}
\email[]{}

\author[]{Jingjing Li}
\affiliation{Purple Mountain Observatory, Chinese Academy of Sciences, Nanjing 210008,People’s Republic of China;}
\affiliation{University of Science and Technology of China, Jinzhai Road, Hefei 230026, People’s Republic of China;}
\email[]{}

\author[]{Zhenyu Wu}
\affiliation{National Astronomical Observatories, Chinese Academy of Sciences, 20A Datun Road, Chaoyang District, Beijing 100101, People’s Republic of China;}
\affiliation{School of Astronomy and Space Science, University of Chinese Academy of Sciences, Beijing 101408, People’s Republic of China}
\email[]{}

\author[]{Jianrong Shi}
\affiliation{National Astronomical Observatories, Chinese Academy of Sciences, 20A Datun Road, Chaoyang District, Beijing 100101, People’s Republic of China;}
\email[]{}

\author[]{Ye Xu}
\altaffiliation{Corresponding Author}
\affiliation{Purple Mountain Observatory, Chinese Academy of Sciences, Nanjing 210008,People’s Republic of China;}
\affiliation{University of Science and Technology of China, Jinzhai Road, Hefei 230026, People’s Republic of China;}
\affiliation{Xinjiang Astronomical Observatory, Chinese Academy of Sciences, Urumqi, Xinjiang, 830011, People's Republic of China}
\email[show]{xuye@pmo.ac.cn}

\begin{abstract}

The search for massive compact objects such as black holes (BHs) or neutron stars (NSs) in star clusters is longstanding challenge in astrophysics, yet no definitive identification of such compact objects in young clusters has been reported. In this Letter, we construct a sample of high-priority candidates for follow-up investigation. By cross-matching the cluster catalog with the \gaia\ Non-Single-Star database and estimating companion masses, we identify 13 systems with companion masses exceeding \(1.4\,\msol\), three of which have masses above \(2.1\,\msol\). Applying our newly developed Astrometric Mass-Ratio Function (AMRF) classification based on cluster-specific isochrones, we identify three systems whose companions cannot readily be explained by a single main-sequence star or an unresolved close binary. These systems have inferred companion masses exceeding 2.1~$\msol$, suggesting that they may be high-mass compact-object candidates. Joint \gaia\ and radial velocity orbit fitting for a comparison subset yields companion masses broadly consistent with the astrometric estimates, supporting the use of our framework for target prioritization. These unconfirmed systems are valuable targets for multi-epoch spectroscopy, high-resolution imaging, and future \gaia\ epoch astrometry.

\end{abstract}

\keywords{\uat{Astrometric binary stars}{79} --- \uat{Compact binary stars}{283} --- \uat{Compact objects}{288} --- \uat{Open clusters and associations}{1160}}


\section{Introduction} \label{sect:intro}
Star clusters are fundamental sites of star formation and early dynamical evolution, and serve as natural laboratories for studying multiple star systems and compact objects. Most stars form in cluster or association environments, where massive stars rapidly evolve into neutron stars (NSs) or black holes (BHs). Dynamical processes within clusters, including mass segregation, mass transfer, and binary interactions, can significantly alter the composition and orbital architecture of binary systems, thereby efficiently promoting the formation or reassembly of binaries hosting compact objects \citep{Duchene+2013, Offner+2023}. Therefore, searching for dark companions in star clusters to identify compact objects offers not only unique dynamical advantages but also higher feasibility and detection efficiency compared to the diffuse field-star environment.

However, to date, no BH or NS has been directly and unambiguously identified in a young cluster. Indirect evidence, primarily through kinematic backtracking, suggests that certain pulsars may have originated in young cluster environments \citep[e.g.,][]{Liu+2025, Zhou+2025, Zhang+2026}. The Non-Single Star (NSS) catalog of \gaia\ DR3 \citep{GaiaNSS+2023} provides an unprecedented observational opportunity for the systematic investigation of dark companions in clusters. Although numerous compact-object candidates, such as \gaia\ BH1, BH2, and BH3, have been identified using the \gaia\ NSS catalog \citep{El-Badry+2023a, El-Badry+2023b, El-Badry+2024}, existing work placed these binary system in isolated environments.

As natural laboratories for multiple-star-system studies, cluster members are born from the same molecular cloud and therefore share highly consistent ages, distances, metallicities, and kinematic properties \citep{Lada+2003, Cantat-Gaudin+2022}. For single star, properties such as mass and evolutionary stage are usually estimated from photometric or spectroscopic data of individual objects \citep[e.g.,][]{Kurtz+2022,Vines+2022,Kordopatis+2023}. Cluster isochrones provide mass--luminosity relations (MLRs) and evolutionary-stage constraints that improve the parameter estimation accuracy for member stars \citep[e.g.,][]{Cantat-Gaudin+2022, Hunt+2023, Hunt+2024}.

In this work, we cross-match the \gaia\ DR3 NSS catalog with the cluster membership catalog of \citet{Hunt+2024} to construct a sample of astrometric binaries in clusters. Cluster parameters are directly incorporated into the classification of dark companions. Specifically, the MLR of the visible star is derived from cluster isochrones, which is constrained by the cluster age, metallicity, distance, and extinction. This approach enables a direct assessment of whether the dark companion can be explained by a single star or an unresolved multiple system. In contrast to methods that adopt the average MLR in the Solar neighborhood \citep{Pecaut+2013,Shahaf+2019,Shahaf+2023}, our new framework moves beyond uniform criteria to deliver a more physically tailored classification, adapted to the actual conditions of individual clusters and providing a cluster-specific ranking of candidates for follow-up.

Furthermore, for stars with multi-epoch radial velocity (RV) observations, we perform joint \gaia\ astrometric and RV orbit fitting \citep{El-Badry+2023a,El-Badry+2023b} to assess the consistency of the astrometric mass estimates for a comparison subset. By combining cluster membership, cluster-specific stellar evolution models, and orbital information, this work aims to prioritize systems for follow-up observations that can determine whether their unseen companions are compact objects.

\section{Data and Analysis} \label{sec:data}
\subsection{Cluster Members and \gaia\ Astrometric Binaries}

We first cross-matched the cluster membership catalog of \citet{Hunt+2024} with the \gaia\ NSS catalog, yielding 1656 \gaia\ astrometric binary candidates distributed across 890 star clusters. The Thiele--Innes parameters provided in the \gaia\ NSS catalog were then converted into Campbell orbital elements using the \texttt{nsstool} package \citep{Halbwachs+2023}. Among these, 677 sources had complete sets of orbital parameters for analysis.

In the catalog of \citet{Hunt+2024}, stellar parameters for most cluster members were derived through PARSEC \citep{Bressan+2012} isochrone fitting. A total of 655 sources had both mass estimates and binary orbital elements. In this work, the parallax and proper motion in the \gaia\ NSS catalog were employed for analysis. 
As preliminary step before the NSS-based analysis, the cluster compatibility of each star was reassessed using updated astrometric solutions (see details in Appendix~\ref{subsect:cluster}). A total of 582 stars were determined to meet the compatibility criteria, forming the final sample for our analysis.
The binary classifications in the \gaia\ NSS catalog for these sources include \texttt{AstroSpectroSB1}, \texttt{Orbital}, \texttt{OrbitalAlternative}, \texttt{OrbitalTargetedSearch}, and \texttt{OrbitalTargetedSearchValidated}.

\subsection{Companion Mass Estimates}
\label{subsect:m2_est}

For each system with a \gaia\ astrometric orbital solution, the astrometric mass function $f_M$ can be computed from the photocentric semi-major axis ($a_0$), period ($P$), and parallax ($\varpi$):
\begin{equation}\label{eq:fM}
  f_M = (a_0/\varpi)^3 (1\;\mathrm{yr}/P)^2 \ \msol.
\end{equation}

Assuming the companion is dark, the dynamical lower-limit mass of the dark companion ($M_2$) can be derived from the mass of the visible star ($M_1$) via:
\begin{equation}\label{eq:M2}
  f_M = M_2 \left(\frac{M_2}{M_1 + M_2}\right)^{\!2}.
\end{equation}

Before estimating the companion masses, we performed multi-band spectral energy distribution (SED) fitting to assess the reliability of the mass derived from cluster isochrone fitting by \citet{Hunt+2024}. Cross-matching our sample with LAMOST DR13\footnote{\url{https://www.lamost.org/dr13/v1.0/}} \citep{Cui+2012, Luo+2015} yields 78 sources with available spectra for SED analysis (see Appendix~\ref{subsect:sed} for details). The SED-derived masses are in good agreement with the isochrone-fitting results of \citet{Hunt+2024}. Given this consistency and the limited cross-match size, we adopt the visible star masses ($M_{1}$) from \citet{Hunt+2024} for our dark-companion analysis to maximize the sample of compact candidates.

Among the 582 sources, the dark companion mass estimates are distributed as follows: three sources with $M_2 > 2.1\,\msol$; 10 sources with $1.4\,\msol < M_2 \leq 2.1\,\msol$; 141 sources with $0.7\,\msol < M_2 \leq 1.4\,\msol$; and 428 sources with $M_2 \leq 0.7\,\msol$. Table~\ref{tab:m2_geq_1_4} lists the 13 massive sample with companion mass estimates exceeding $1.4\,\msol$. They have intermediate-to-long orbital periods of 457–1187 days, with a median of approximately 705 days. Their eccentricities span a broad range, $e=0.13$–0.67, and no single orbital morphology characterizes the entire sample. 

For these 13 systems, the visible components are relatively massive, spanning $M_{1} = 2.00-$5.76~$\msol$, while the inferred companion masses range from 1.42 to 3.05~$\msol$, with a median of $1.80\,\msol$. The corresponding mass ratios ($q = M_2 / M_1$) range from 0.34 to 0.94, with a median of 0.58. 

The host clusters of these 13 systems span ages from approximately 4~Myr to 760~Myr, though the sample is weighted toward young populations. Eight of the 13 systems belong to clusters younger than 100~Myr, and seven are younger than 30~Myr. The presence of a genuine compact remnant in young clusters would require an exceptionally massive progenitor and rapid post-main-sequence evolution. These systems are consequently of particular interest, but they also require especially careful validation of the cluster membership, age, astrometric orbit, and possible contribution from unresolved luminous companions.

\begin{figure}[htbp]
    \centering
    \includegraphics[width=0.6\linewidth]{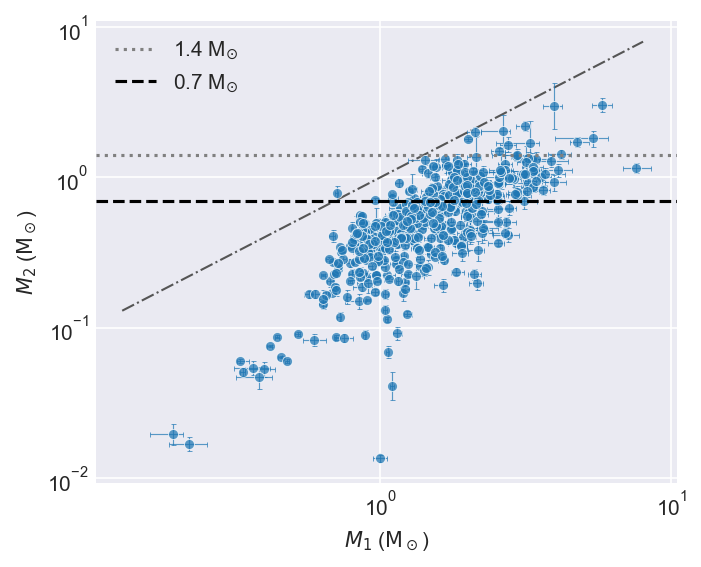}
    \caption{Mass relations. Visible star mass $M_1$ versus dark companion mass $M_2$. The horizontal black dashed and gray dotted lines mark $M_2 = 0.7\,\msol$ and $M_2 = 1.4\,\msol$, respectively. The red dash-dotted line indicates the 1:1 relation.}
    \label{fig:m1_vs_m2}
\end{figure}

Figure~\ref{fig:m1_vs_m2} presents the mass correlation between the visible star and the dark companion. For companions with $M_2 \leq 0.7\,\msol$, the median and mean mass ratios are 0.36 and 0.35, respectively; for the $0.7\,\msol < M_2 \leq 1.4\,\msol$ group, the corresponding values are 0.42 and 0.46. This suggests that systems with more massive dark companions tend to occupy high mass-ratio regimes. Although the current sample with $M_2 > 1.4\,\msol$ remains too limited for robust statistical inference, its median ratio of 0.58 is notably higher than those of the other groups.

\begin{table}[htbp]
    \centering
    \small
    \caption{Sample of Systems with Dark Companion Mass
    $M_2 > 1.4\,\msol$ \label{tab:m2_geq_1_4}}
    \setlength{\tabcolsep}{2pt} 
    \begin{tabular}{ccccccccc}
    \hline \hline
    Gaia ID & NSS Type & $M_{1}$ & $M_{2}$ & $P$ & $e$ & parallax & Cluster & Log $t$ \\
    &  & ($\msol$) & ($\msol$)  & (day) & & (mas) & &   \\ \hline
    5328580248426543744 & Orbital & $5.76^{+0.51}_{-0.43}$ & $3.05^{+0.35}_{-0.32}$ & $842.1 \pm 47.9$ & $0.57 \pm 0.09$ & $0.50 \pm 0.02$ & Muzzio 1 & $7.11^{+0.22}_{-0.25}$ \\
    5334365427630804608 & Orbital & $3.95^{+0.26}_{-0.31}$ & $3.00^{+1.27}_{-0.86}$ & $1186.8 \pm 256.6$ & $0.62 \pm 0.12$ & $0.28 \pm 0.02$ & UBC 1500 & $7.33^{+0.17}_{-0.21}$ \\
    5932802153199174912 & Orbital & $3.14^{+0.11}_{-0.22}$ & $2.21^{+0.18}_{-0.16}$ & $663.2 \pm 15.2$ & $0.20 \pm 0.06$ & $0.53 \pm 0.02$ & NGC 6031 & $8.03^{+0.19}_{-0.18}$ \\ \hline
    4065974380116009472 & Orbital & $2.64^{+0.15}_{-0.43}$ & $2.03^{+0.55}_{-0.46}$ & $915.6 \pm 47.9$ & $0.67 \pm 0.14$ & $0.80 \pm 0.02$ & NGC 6530 & $6.59^{+0.08}_{-0.14}$ \\
    1857469799576340864 & Orbital & $2.11^{+0.11}_{-0.13}$ & $1.99^{+0.16}_{-0.14}$ & $908.6 \pm 38.7$ & $0.13 \pm 0.08$ & $0.88 \pm 0.03$ & NGC 6940 & $8.89^{+0.19}_{-0.16}$ \\
    512069335595731712 & AstroSpectroSB1 & $5.37^{+0.71}_{-1.38}$ & $1.83^{+0.21}_{-0.22}$ & $821.3 \pm 1.0$ & $0.39 \pm 0.01$ & $0.94 \pm 0.01$ & HSC 1039 & $7.94^{+0.20}_{-0.20}$ \\
    5339426861963121280 & Orbital & $2.00^{+0.08}_{-0.05}$ & $1.80^{+0.05}_{-0.05}$ & $647.2 \pm 2.5$ & $0.30 \pm 0.02$ & $2.04 \pm 0.02$ & NGC 3532 & $8.38^{+0.28}_{-0.22}$ \\
    2061381034190368256 & Orbital & $4.73^{+0.46}_{-0.25}$ & $1.72^{+0.13}_{-0.12}$ & $556.2 \pm 6.6$ & $0.23 \pm 0.07$ & $0.57 \pm 0.01$ & vdBergh 130 & $6.66^{+0.11}_{-0.16}$ \\
    2058573117692962816 & AstroSpectroSB1 & $3.27^{+0.24}_{-0.77}$ & $1.70^{+0.68}_{-0.50}$ & $1135.3 \pm 222.5$ & $0.19 \pm 0.12$ & $0.42 \pm 0.01$ & Gulliver 38 & $8.59^{+0.20}_{-0.21}$ \\
    6056465089314101376 & OrbitalAlternative & $2.75^{+0.17}_{-0.17}$ & $1.65^{+0.22}_{-0.18}$ & $705.1 \pm 8.6$ & $0.21 \pm 0.05$ & $0.47 \pm 0.04$ & NGC 4755 & $7.37^{+0.15}_{-0.21}$ \\
    534269879805569280 & Orbital & $2.57^{+0.12}_{-0.16}$ & $1.50^{+0.08}_{-0.08}$ & $456.7 \pm 1.9$ & $0.31 \pm 0.04$ & $1.16 \pm 0.02$ & Collinder 463 & $8.17^{+0.24}_{-0.15}$ \\
    2062361145712093568 & Orbital & $4.17^{+0.36}_{-0.39}$ & $1.43^{+0.10}_{-0.09}$ & $524.9 \pm 3.6$ & $0.19 \pm 0.04$ & $1.01 \pm 0.01$ & Collinder 419 & $6.72^{+0.13}_{-0.17}$ \\
    2067032867532605056 & Orbital & $2.94^{+0.35}_{-0.52}$ & $1.42^{+0.12}_{-0.13}$ & $565.8 \pm 2.7$ & $0.20 \pm 0.03$ & $1.19 \pm 0.01$ & UPK 126 & $6.61^{+0.12}_{-0.16}$ \\ \hline
    \end{tabular}
\end{table}

\section{Companion Classification Based on the AMRF}
\subsection{Classification Criteria}

To further constrain the true physical nature of the dark companions, we adopted the Astrometric Mass-Ratio Function AMRF, $\mathcal{A}$) classification method of \citet{Shahaf+2019, Shahaf+2023}. This method classifies astrometric binaries into three classes:
\begin{itemize}
  \item \textbf{Class~I}: The companion is most likely a single main-sequence star.
  \item \textbf{Class~II}: The companion is most likely an unresolved close main-sequence binary rather than a single main-sequence star.
  \item \textbf{Class~III}: The companion is most likely a compact object. It can be neither a single main-sequence star nor a close main-sequence binary.
\end{itemize}

We applied the quality cuts of \citet{Halbwachs+2023} to ensure the reliability of the orbital solutions:
\begin{itemize}
  \item $\Delta e < 0.079 \ln(P/\mathrm{day}) - 0.244$;
  \item $\varpi/\Delta\varpi > 20000\,(P/\mathrm{day})^{-1}$; 
  \item $a_0/\Delta a_0 > 158\,(P/\mathrm{day})^{-1/2}$; and
  \item \texttt{goodness\_of\_fit} $<$ 5.
\end{itemize}

In addition, following \citet{Shahaf+2023}, a constraint on the uncertainties of the Thiele--Innes parameters provided by \gaia\ DR3 was imposed:
\begin{equation}\label{eq:sigmaTI}
  \sigma^2_{\mathrm{TI}}
  = \left(\frac{\Delta A}{A}\right)^{\!2}
  + \left(\frac{\Delta B}{B}\right)^{\!2}
  + \left(\frac{\Delta F}{F}\right)^{\!2}
  + \left(\frac{\Delta G}{G}\right)^{\!2}
  \leq 36.
\end{equation}
Of the initial 582 sources, 367 met the above selection criteria.

We extended the AMRF framework of \citet{Shahaf+2019,Shahaf+2023}  by replacing their single MLR with cluster-specific isochrones. Rather than relying on fixed boundaries \citep{Shahaf+2019,Shahaf+2023}, we adopted PARSEC isochrones \citep{Bressan+2012} matched to the age and metallicity of each host cluster to derive the appropriate MLR and compute dedicated AMRF boundaries per cluster. This approach avoids the errors introduced by projecting members of different ages and metallicities onto a single MLR, and naturally accommodates systems with $M_1 > 2\,\msol$.

To properly propagate observational uncertainties, we performed Monte Carlo sampling over the visible-stellar mass, cluster age, and orbital parameters. For each target, the classification frequency across independent realizations was recorded, and these frequencies were converted into membership probabilities $\mathrm{Pr_{I}}$, $\mathrm{Pr_{II}}$, and $\mathrm{Pr_{III}}$ for Classes~I, II and III, respectively.

\subsection{Classification Results}

To select high-confidence compact-object candidates, stringent cuts were imposed: 1) more than 80\% of Monte Carlo realizations were required to be effectively covered by the host cluster's PARSEC evolutionary model, and 2) the cluster membership probability (Pr$_{\rm Member}$) of each target was required to exceed 50\%. The final selection yielded three high-priority Class~III candidates with $\mathrm{Pr_{III}} > 50\%$, 77~high-priority Class~II candidates with $\mathrm{Pr_{II}} > 50\%$, and 221~high-priority Class~I candidates with $\mathrm{Pr_{I}} > 50\%$. Of the 13 systems with $M_2 > 1.4\,\msol$ mentioned in Table~\ref{tab:m2_geq_1_4}, four meet the above selection criteria. Among these, three are classified as high-priority Class~III candidates, and one remains ambiguous, with probabilities of 49\% for Class~II and 46\% for Class~III. 

\begin{figure}[htbp]
\centering
    \begin{subfigure}[]{0.45\linewidth}
        \centering
        \includegraphics[width=\linewidth]{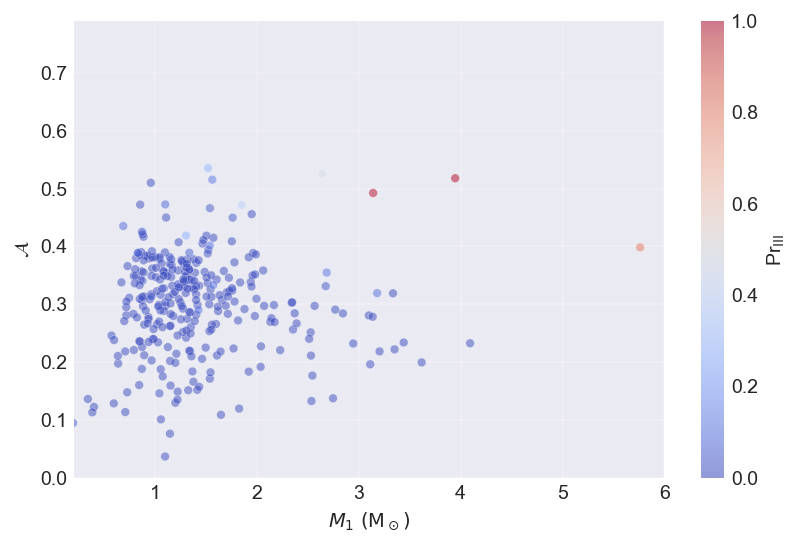}
        \caption{}
        \label{fig:amrf_all}
    \end{subfigure}
    \begin{subfigure}[]{0.45\linewidth}
        \centering
        \includegraphics[width=\linewidth]{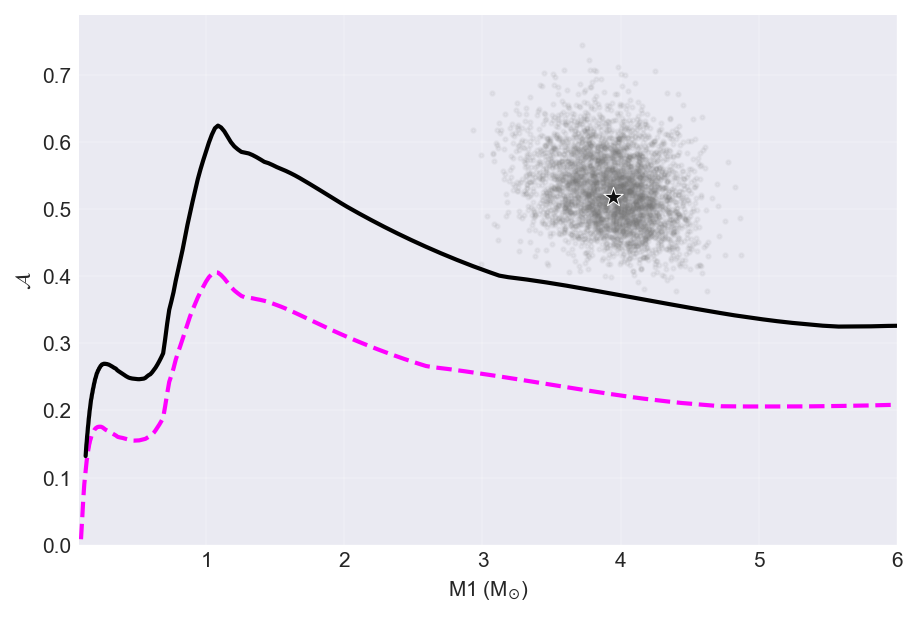}
        \caption{}
        \label{fig:amrf_ex}
    \end{subfigure}
    \caption{AMRF--$M_1$ distribution. (a): Distribution of all sources in the AMRF--$M_1$ plane. The color bar encodes the probability of Class~III classification. (b): Example of AMRF--$M_1$ distribution for a single source (\gaia\ DR3\,5334365427630804608). The black star marks the stellar position, and gray scatter points are Monte Carlo realizations. The solid black line denotes the boundary between Class~II and~III, the magenta dashed line denotes the boundary between Class~I and~II.}
    \label{fig:amrf_parsec}
\end{figure}

Table~\ref{tab:amrf} lists the basic properties of the three high-priority Class~III systems, whose dark companions cannot readily be explained by a single main-sequence star or an unresolved close binary, while compact objects and more complex higher-order multiples are possible interpretations.
All three high-priority Class III systems have companion masses exceeding 2.1~$\msol$, suggesting that they are promising high-mass compact-object candidates.
Figure~\ref{fig:amrf_all} displays the distribution of the full sample in the AMRF--$M_1$ parameter space, color-coded by the probability of Class~III membership. An example classification is highlighted in Figure~\ref{fig:amrf_ex}, demonstrating a high-priority Class~III system.

Figure~\ref{fig:clusters} in Appendix~\ref{subsect:cluster} displays the distribution of these sources within their respective host clusters. As shown in Figure~\ref{fig:clusters}, the three massive compact-object candidates are located in the central regions of their host clusters, yet lie at the periphery in proper-motion space, with parallaxes shifted away from the core histogram. 

\begin{table}[htbp]
    \centering
    \caption{High-Priority Class~III Candidates  \label{tab:amrf}}
    \begin{tabular}{ccccccccc}
    \hline \hline
    Source ID & $M_2$& $\mathcal{A}$ & Pr$_{\rm I}$ & Pr$_{\rm II}$  & Pr$_{\rm III}$ & Cluster & Pr$_{\rm Member}$ & \texttt{goodness\_of\_fit} \\ 
    & ($\msol$) & & (\%) & (\%) & (\%) & & (\%) & \\ \hline
    5328580248426543744 & $3.05^{+0.35}_{-0.32}$ & 0.40 & 0.00 & 17.32 & 82.68 & Muzzio 1 & 96.07 & 0.40 \\
    5334365427630804608 & $3.00^{+1.27}_{-0.86}$ & 0.52 & 0.00 & 0.02 & 99.98 & UBC 1500 & 100.00 & -2.57 \\
    5932802153199174912 & $2.21^{+0.18}_{-0.16}$ & 0.49 & 0.00 & 0.45 & 99.55 & NGC 6031 & 91.05 & -0.89 \\
    \hline
    \end{tabular}
\end{table}

\section{Radial Velocity Validation}

To obtain multi-epoch RVs, we cross-matched the LAMOST \citep{Cui+2012, Luo+2015}, APOGEE \citep{SDSS+2025,Meszaros+2025}, and GALAH \citep{Buder+2025} surveys, supplemented by the multi-epoch RV dataset for nearby clusters published by \citet{Mermilliod+2009}. For the 582 sources in our sample, this yielded 37 stars with multi-epoch RVs, 31 of which had at least three epochs. Unfortunately, for the 13 massive star candidates in Table~\ref{tab:m2_geq_1_4} and the three high-priority Class III candidates in Table~\ref{tab:amrf}, our cross-match returned no multi-epoch RV measurements from these datasets.

Following the dynamical modeling approach of \citet{El-Badry+2023a,El-Badry+2023b}, joint \gaia+RV orbit fitting was performed for these 31 sources with three or more epochs. MCMC sampling and posterior parameter estimation were carried out using the \texttt{emcee} package \citep{Foreman+2013}. Figure~\ref{fig:mcmc_summary} in Appendix~\ref{subsect:mcmc_fit} presents the distributions of the multi-dimensional quality metrics after joint orbit fitting. Adopting the classification criteria defined in Appendix~\ref{subsect:mcmc_fit}, we rated 14 sources as having ``good'' joint-fit quality and the remaining 17 as ``poor'', owing to insufficient epoch coverage or waveform distortion. Table~\ref{tab:mcmc_good} in Appendix~\ref{subsect:mcmc_fit} summarizes the 14 joint orbital solutions that met the good criteria.

\begin{figure}[htbp]
    \centering
    \includegraphics[width=0.6\linewidth]{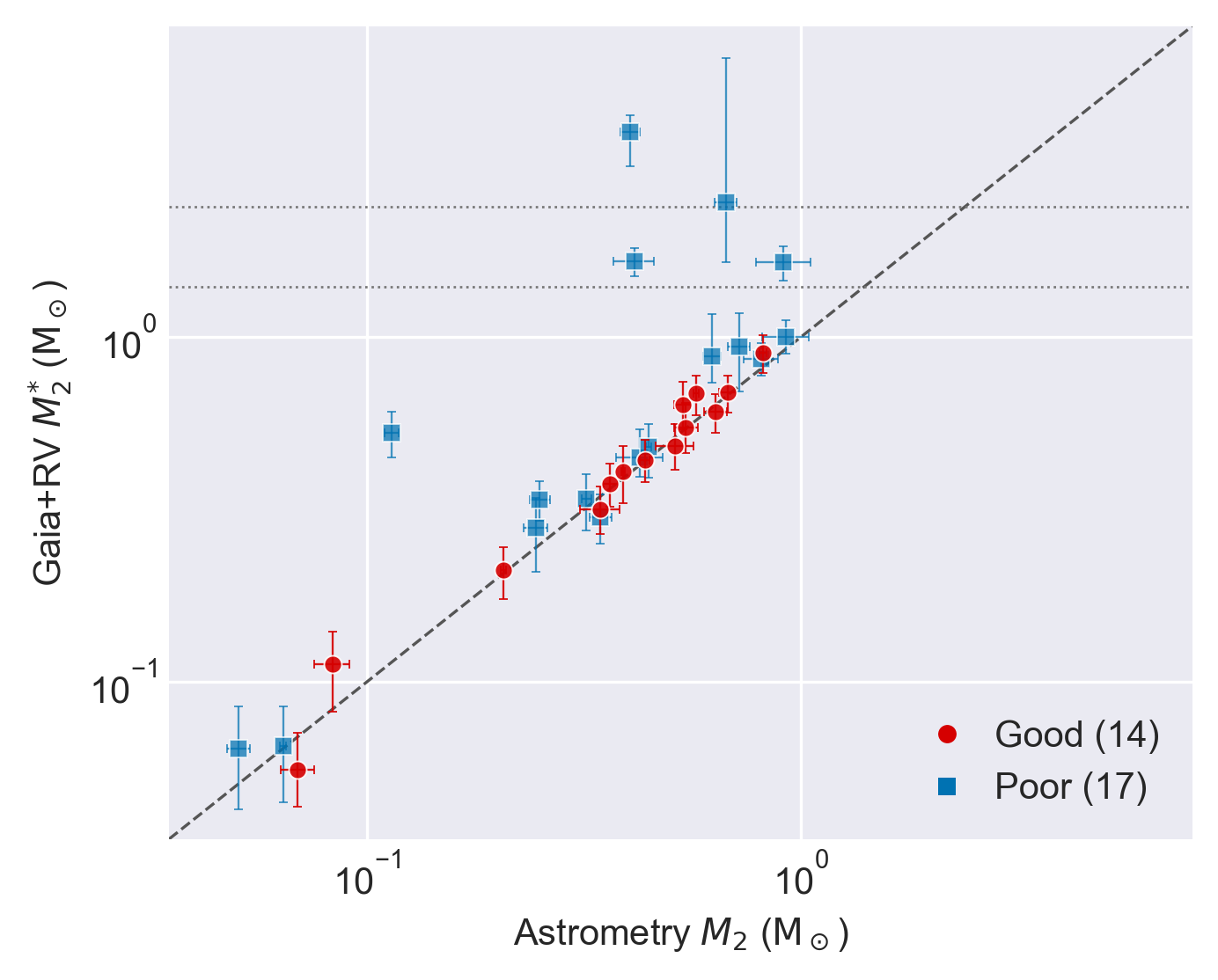}
    \caption{Companion masses estimated from astrometry alone ($M_{2}$, Section \ref{subsect:m2_est}) versus those from the joint Gaia+RV orbit fit ($M_{2}^{*}$). Red points and blue squares denote good and poor fits, respectively. Horizontal dotted lines mark $1.4\,\msol$ and $2.1\,\msol$. The diagonal dashed line is the 1:1 relation.}
    \label{fig:m2_astro_gaia_rv}
\end{figure}

Figure~\ref{fig:m2_astro_gaia_rv} compares the unknown companion masses derived from astrometry ($M_{2}$, Section~\ref{subsect:m2_est}) with those obtained from the joint \gaia+RV orbit fitting ($M_{2}^{*}$). For systems with good quality, the masses derived from the two methods are in excellent agreement. Although four systems in the multi-epoch RV sample have joint-fit masses $M_{2}^{*} > 1.4\,\msol$, the astrometric masses ($M_{2}$) does not exceed $1.4\,\msol$. Moreover, their joint \gaia+RV fits are rated as poor, preventing tighter dynamical constraints on their companion masses at this stage.  
Although none of the 13 massive stellar candidates or the three high-priority Class III candidates are included in Figure~\ref{fig:m2_astro_gaia_rv}, the agreement exhibited by the good-quality comparison systems provides an internal consistency check on the astrometry-only mass-estimation procedure over the parameter range sampled by the RV data.

\section{Summary and Conclusions}

Stars predominantly form in star clusters, where massive stars, the progenitors of BHs or NSs, are expected to reside during their early evolution. However, most clusters are likely to be disrupted before their massive members undergo core collapse and produce compact remnants. Furthermore, even when BHs or NSs form within clusters, they may be ejected by natal kicks imparted during supernova explosions. As a result, BHs or NSs in young clusters remain rare, their unambiguous identification would offer unique insights into BH/NS formation mechanisms and stellar evolution.

In this Letter, we search for compact companions in young star clusters by cross-matching the cluster catalog of \citet{Hunt+2024} with the \gaia\ DR3 NSS database, yielding a sample of 1656 astrometric binary candidates associated with 890 star clusters. Among these, 582 systems have visible-star masses and complete orbital solutions, and additionally met the cluster compatibility criteria. 13 systems have inferred companion masses $M_2 > 1.4\,\msol$, including three with $M_2 > 2.1\,\msol$. Using the PARSEC isochrones of each host cluster, we extend the AMRF framework of \citet{Shahaf+2019,Shahaf+2023}. Based on the new approach, we find three systems whose companions cannot be explained by a single main-sequence star or an unresolved close binary. These systems have companion masses of $\sim 2.21$--$3.05\,\msol$ and are classified as high-mass compact-object candidates.
These designations are intended to rank targets for follow-up and do not constitute physical identifications.

As a consistency check, we collected multi-epoch RV data and performed joint \textit{Gaia}+RV orbit fitting for 31 systems with sufficient observational coverage. For systems with good fits, the joint-fit masses are broadly consistent with those inferred from the \gaia\ photocentric orbits. Therefore, the 13 objects with $M_2 > 1.4\,\msol$ derived from \gaia\ photocentric orbits define clear targets for precision follow-up observations, particularly including the three massive high-priority Class~III candidates identified via the new AMRF classification. Table~\ref{tab:sum} summarizes the data for all sources with $M_{2} > 1.4\,\msol$, all high-confidence Class III sources, and all sources with a good Gaia+RV fit quality rating.

While current spectroscopic epoch coverage limits immediate joint-solution confirmation for the most massive candidates, these high-priority sources constitute prime targets for urgent follow-up. Future ground-based high-resolution spectroscopy and \textit{Gaia} DR4 epoch data will critically test these systems, providing direct, decisive constraints on compact object formation, evolution, and dynamical retention within cluster environments.

\begin{table}[]
    \centering
    \small
    \setlength{\tabcolsep}{2pt} 
    \caption{Summary of the Samples Used in This Work}
    \begin{tabular}{cccccccc}
    \hline \hline
    Source ID & NSS Type & Cluster & $M_{2}$ & Mass Selected & Class III Candidates & Joint Fit & Fit Quality  \\ \hline
    5328580248426543744 & Orbital & Muzzio 1 & $3.05^{+0.35}_{-0.32}$ & Yes & Yes & No & No \\
    5334365427630804608 & Orbital & UBC 1500 & $3.00^{+1.27}_{-0.86}$ & Yes & Yes & No & No \\
    5932802153199174912 & Orbital & NGC 6031 & $2.21^{+0.18}_{-0.16}$ & Yes & Yes & No & No \\
    \hline
    4065974380116009472 & Orbital & NGC 6530 & $2.03^{+0.55}_{-0.46}$ & Yes  & No & No & No \\
    1857469799576340864 & Orbital & NGC 6940 & $1.99^{+0.16}_{-0.14}$ & Yes  & No & No & No \\
    512069335595731712 & AstroSpectroSB1 & HSC 1039 & $1.83^{+0.21}_{-0.22}$ & Yes  & No & No & No \\
    5339426861963121280 & Orbital & NGC 3532 & $1.80^{+0.05}_{-0.05}$ & Yes  & No & No & No \\
    2061381034190368256 & Orbital & vdBergh 130 & $1.72^{+0.13}_{-0.12}$ & Yes  & No & No & No \\
    2058573117692962816 & AstroSpectroSB1 & Gulliver 38 & $1.70^{+0.68}_{-0.50}$ & Yes  & No & No & No \\
    6056465089314101376 & OrbitalAlternative & NGC 4755 & $1.65^{+0.22}_{-0.18}$ & Yes  & No & No & No \\
    534269879805569280 & Orbital & Collinder 463 & $1.50^{+0.08}_{-0.08}$ & Yes  & No & No & No \\
    2062361145712093568 & Orbital & Collinder 419 & $1.43^{+0.10}_{-0.09}$ & Yes  & No & No & No \\
    2067032867532605056 & Orbital & UPK 126 & $1.42^{+0.12}_{-0.13}$ & Yes  & No & No & No \\ \hline
    661148268907314432 & AstroSpectroSB1 & NGC 2632 & $0.82^{+0.02}_{-0.01}$ & No  & No & Yes & Good \\
    2275682653647151232 & AstroSpectroSB1 & HSC 976 & $0.68^{+0.03}_{-0.03}$ & No  & No & Yes & Good \\
    1987734091775898752 & Orbital & UPK 167 & $0.64^{+0.04}_{-0.04}$ & No  & No & Yes & Good \\
    3953951879155969920 & AstroSpectroSB1 & Melotte 111 & $0.57^{+0.01}_{-0.01}$ & No  & No & Yes & Good \\
    598974799070364928 & Orbital & NGC 2682 & $0.54^{+0.04}_{-0.03}$ & No  & No & Yes & Good \\
    4040823292139528320 & Orbital & NGC 6475 & $0.54^{+0.02}_{-0.02}$ & No  & No & Yes & Good \\
    598960058742607744 & Orbital & NGC 2682 & $0.51^{+0.05}_{-0.05}$ & No  & No & Yes & Good \\
    65090680344356992 & Orbital & Melotte 22 & $0.44^{+0.01}_{-0.01}$ & No  & No & Yes & Good \\
    3314151251273992832 & Orbital & Melotte 25 & $0.39^{+0.00}_{-0.00}$ & No  & No & Yes & Good \\
    454771929245197312 & Orbital & CWNU 1076 & $0.36^{+0.01}_{-0.01}$ & No  & No & Yes & Good \\
    604969783142095744 & Orbital & NGC 2682 & $0.35^{+0.04}_{-0.03}$ & No  & No & Yes & Good \\
    4572777707832147712 & Orbital & HSC 453 & $0.21^{+0.00}_{-0.00}$ & No  & No & Yes & Good \\
    6235684556882813184 & Orbital & CWNU 1143 & $0.08^{+0.01}_{-0.01}$ & No  & No & Yes & Good \\
    1607476280298633984 & Orbital & HSC 759 & $0.07^{+0.01}_{-0.01}$ & No  & No & Yes & Good \\ \hline
    \end{tabular}
    \label{tab:sum}
\end{table}

\begin{acknowledgments}
We thank the referee for the careful reading of our manuscript and for constructive comments. This work was funded by the NSFC Grands Nos. 12503071, 12403077, 12403041, and 11933011; the National Key R\&D Program of China (grant No.2024YFA1611504); the National SKA Program of China (grant No. 2022SKA0120103); the Xinjiang Talent Development Fund (No. XJRC-2025-KJ-YJ-CXPT-180); and the Key Laboratory for Radio Astronomy. 

This work has made use of data from the European Space Agency (ESA) mission \emph{Gaia} (\url{www.cosmos.esa.int/gaia}), processed by the Gaia Data Processing and Analysis Consortium (DPAC, \url{www.cosmos.esa.int/web/gaia/dpac/consortium}). 

This work made use of the data from LAMOST (Large Sky Area Multi-Object Fiber Spectroscopic Telescope, also known as the Guoshoujing Telescope) (\url{https://cstr.cn/31118.02.LAMOST}). LAMOST is a Chinese national mega-science facility, operated by National Astronomical Observatories, Chinese Academy of Sciences.

SDSS is managed by the Astrophysical Research Consortium for the Participating Institutions of the SDSS Collaboration, including Caltech, The Carnegie Institution for Science, Chilean National Time Allocation Committee (CNTAC) ratified researchers, The Flatiron Institute, the Gotham Participation Group, Harvard University, Heidelberg University, The Johns Hopkins University, L'Ecole polytechnique f{\' e}d{\' e}rale de Lausanne (EPFL), Leibniz-Institut f{$\rm \ddot u$}r Astrophysik Potsdam (AIP), Max-Planck-Institut f{$\rm \ddot u$}r Astronomie (MPIA Heidelberg), Max-Planck-Institut f{$\rm \ddot u$}r Extraterrestrische Physik (MPE), Nanjing University, National Astronomical Observatories of China (NAOC), New Mexico State University, The Ohio State University, Pennsylvania State University, Smithsonian Astrophysical Observatory, Space Telescope Science Institute (STScI), the Stellar Astrophysics Participation Group, Universidad Nacional Aut{\' o}noma de M{\' e}xico, University of Arizona, University of Colorado Boulder, University of Illinois at Urbana-Champaign, University of Toronto, University of Utah, University of Virginia, Yale University, and Yunnan University.

This work made use of the Fourth Data Release of the GALAH Survey (Buder et al. 2021). The GALAH Survey is based on data acquired through the Australian Astronomical Observatory, under programs: A/2013B/13 (The GALAH pilot survey); A/2014A/25, A/2015A/19, A2017A/18 (The GALAH survey phase 1); A2018A/18 (Open clusters with HERMES); A2019A/1 (Hierarchical star formation in Ori OB1); A2019A/15, A/2020B/23, R/2022B/5, R/2023A/4, R2023B/5 (The GALAH survey phase 2); A/2015B/19, A/2016A/22, A/2016B/10, A/2017B/16, A/2018B/15 (The HERMES-TESS program); A/2015A/3, A/2015B/1, A/2015B/19, A/2016A/22, A/2016B/12, A/2017A/14, A/2020B/14 (The HERMES K2-follow-up program); R/2022B/02 and A/2023A/09 (Combining asteroseismology and spectroscopy in K2); A/2023A/8 (Resolving the chemical fingerprints of Milky Way mergers); and A/2023B/4 (s-process variations in southern globular clusters).
\end{acknowledgments}

\appendix

\section{Appendix information}

\subsection{Cluster Properties}
\label{subsect:cluster}

To quantitatively assess whether the NSS astrometric solutions remain compatible with the astrometric distributions of their host clusters, we perform a three-dimensional consistency test in proper-motion and parallax space. Our approach is motivated by the three-dimensional ellipsoids used by \citet[][Section 3.2]{Hunt+2024} to characterize the recovery of cluster members, and follows the covariance-weighted Mahalanobis-distance framework of \citet[][Section 3.2]{Medina+2021}. Rather than re-determining cluster membership, we adopt the mean proper motions, mean parallax, corresponding member dispersions, and standard errors of the means reported by \citet{Hunt+2024} as fixed reference quantities.

For each source, we construct the NSS astrometric vector
\begin{equation}
\boldsymbol{x}_{\rm NSS} = (\mu_{\alpha *},\mu_{\delta},\varpi)^{\rm T},
\end{equation}
and define the corresponding cluster center as
\begin{equation}
\boldsymbol{x}_{\rm cl} = (\mu_{\alpha *,{\rm cl}}, \mu_{\delta,{\rm cl}},  \varpi_{\rm cl})^{\rm T}.
\end{equation}
The residual vector is
\begin{equation}
\Delta\boldsymbol{x} = \boldsymbol{x}_{\rm NSS}-\boldsymbol{x}_{\rm cl}.
\end{equation}

We quantify the normalized astrometric offset using the squared Mahalanobis distance,
\begin{equation}
D^2 = \Delta\boldsymbol{x}^{\rm T} \boldsymbol{C}_{\rm tot}^{-1} \Delta\boldsymbol{x},
\end{equation}
where
\begin{equation}
\boldsymbol{C}_{\rm tot} = \boldsymbol{C}_{\rm cluster} + \boldsymbol{C}_{\rm NSS} + \boldsymbol{C}_{\rm center}.
\end{equation}

Because the \citet{Hunt+2024} catalog does not provide cross-parameter correlations for these quantities, both matrices are approximated as diagonal. The cluster-member dispersion and uncertainty of the cluster center are represented by
\begin{equation}
\boldsymbol{C}_{\rm cluster} = {\rm diag} \left(s_{\mu_{\alpha *}}^2, s_{\mu_{\delta}}^2, s_{\varpi}^2\right)
\end{equation}
and
\begin{equation}
\boldsymbol{C}_{\rm center} = {\rm diag}
\left( e_{\mu_{\alpha *,{\rm cl}}}^2, e_{\mu_{\delta,{\rm cl}}}^2, e_{\varpi,{\rm cl}}^2 \right),
\end{equation}
respectively. Here, $s_{\mu_{\alpha *,{\rm cl}}}$, $s_{\mu_{\delta,{\rm cl}}}$, and $s_{\varpi,{\rm cl}}$ are the standard deviations of the cluster-member astrometry, corresponding to \texttt{s\_pmRA}, \texttt{s\_pmDE}, and \texttt{s\_Plx} in the \citet{Hunt+2024} catalog. The quantities $e_{\mu_{\alpha *,{\rm cl}}}$, $e_{\mu_{\delta,{\rm cl}}}$, and $e_{\varpi,{\rm cl}}$ are the standard errors of the corresponding cluster mean parameters, given by \texttt{e\_pmRA}, \texttt{e\_pmDE}, and \texttt{e\_Plx}.

The NSS source covariance matrix is constructed from the formal uncertainties and correlation coefficients of the NSS orbital solution:
\begin{equation}
(\boldsymbol{C}_{\rm NSS})_{ii}=\sigma_i^2, \qquad (\boldsymbol{C}_{\rm NSS})_{ij}=\rho_{ij}\sigma_i\sigma_j,
\end{equation}
where the index $i = 1, 2, 3$ corresponds to $\mu_{\alpha}^{*}$, $\mu_{\delta}$, and $\varpi$, respectively. The correlation coefficients are extracted from the NSS \texttt{corr\_vec} field using the parameter ordering appropriate to each solution type. Thus, the full correlations among parallax and the two proper-motion components are retained.

Under the null hypothesis that the NSS astrometry is drawn from the three-dimensional Gaussian distribution of the host cluster, $D^2$ is expected to approximately follow a chi-square distribution with three degrees of freedom, i.e., $D^2\sim\chi_3^2$.
We therefore calculate the right-tail probability
\begin{equation}
p_{\rm kin} = P(\chi_3^2\geq D^2) = \int_{D^2}^{\infty} \frac{t^{1/2}\exp(-t/2)}{\sqrt{2\pi}}\,{\rm d}t.
\end{equation}
Here, $p_{\rm kin}$ is a statistical measure of astrometric compatibility, rather than a re-estimated membership probability. 

Under the three-dimensional Gaussian approximation, $p_{\rm kin}=0.05$ corresponds to the boundary of the source-specific 95\% acceptance ellipsoid. We regard sources with $p_{\rm kin}\geq0.05$ as astrometrically compatible with their host clusters, while sources with $p_{\rm kin}<0.05$ are flagged as showing astrometric tension. Finally, a total of 582 stars met the criterion.

\begin{figure}
    \centering
    \includegraphics[width=0.45\linewidth]{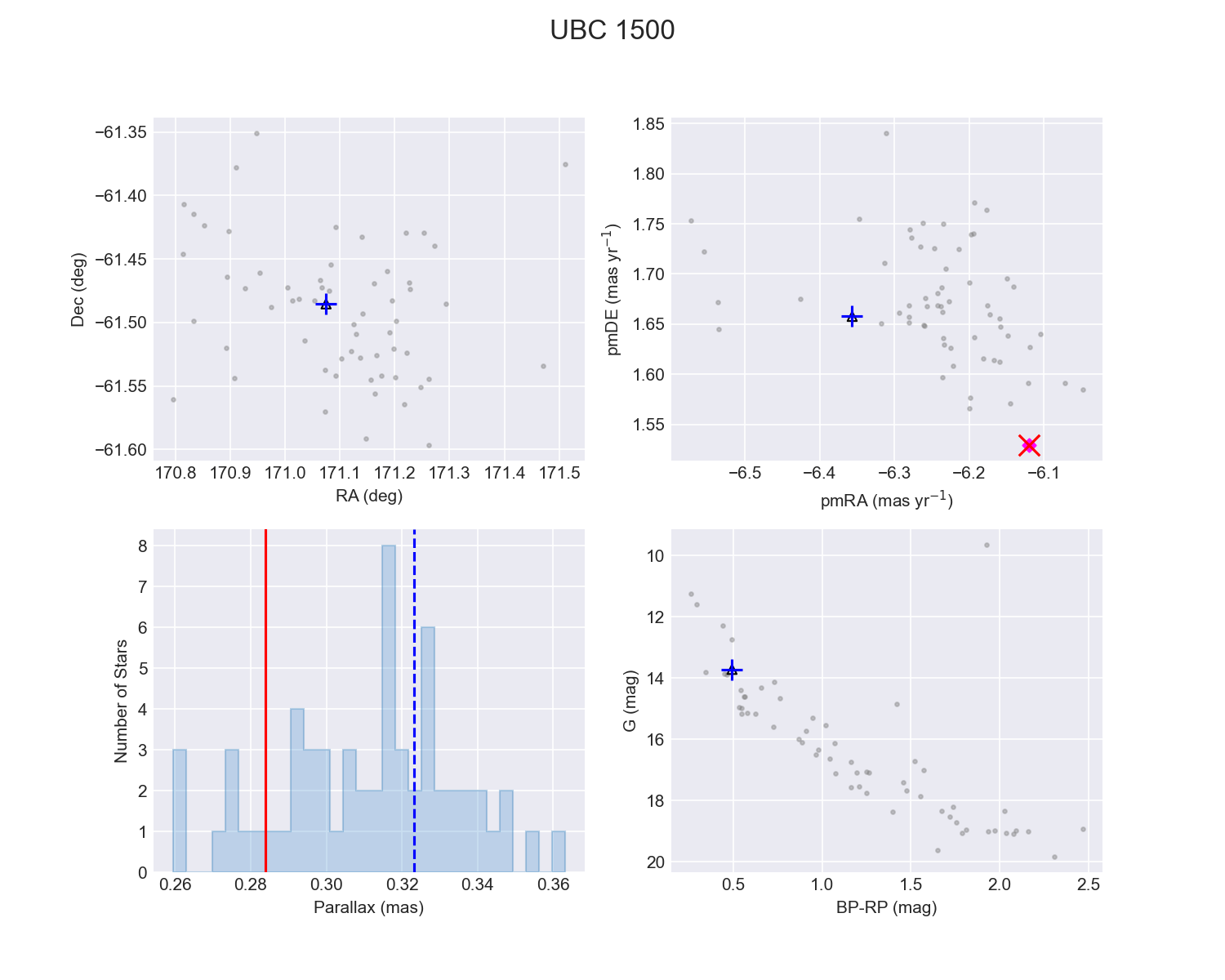}
    \includegraphics[width=0.45\linewidth]{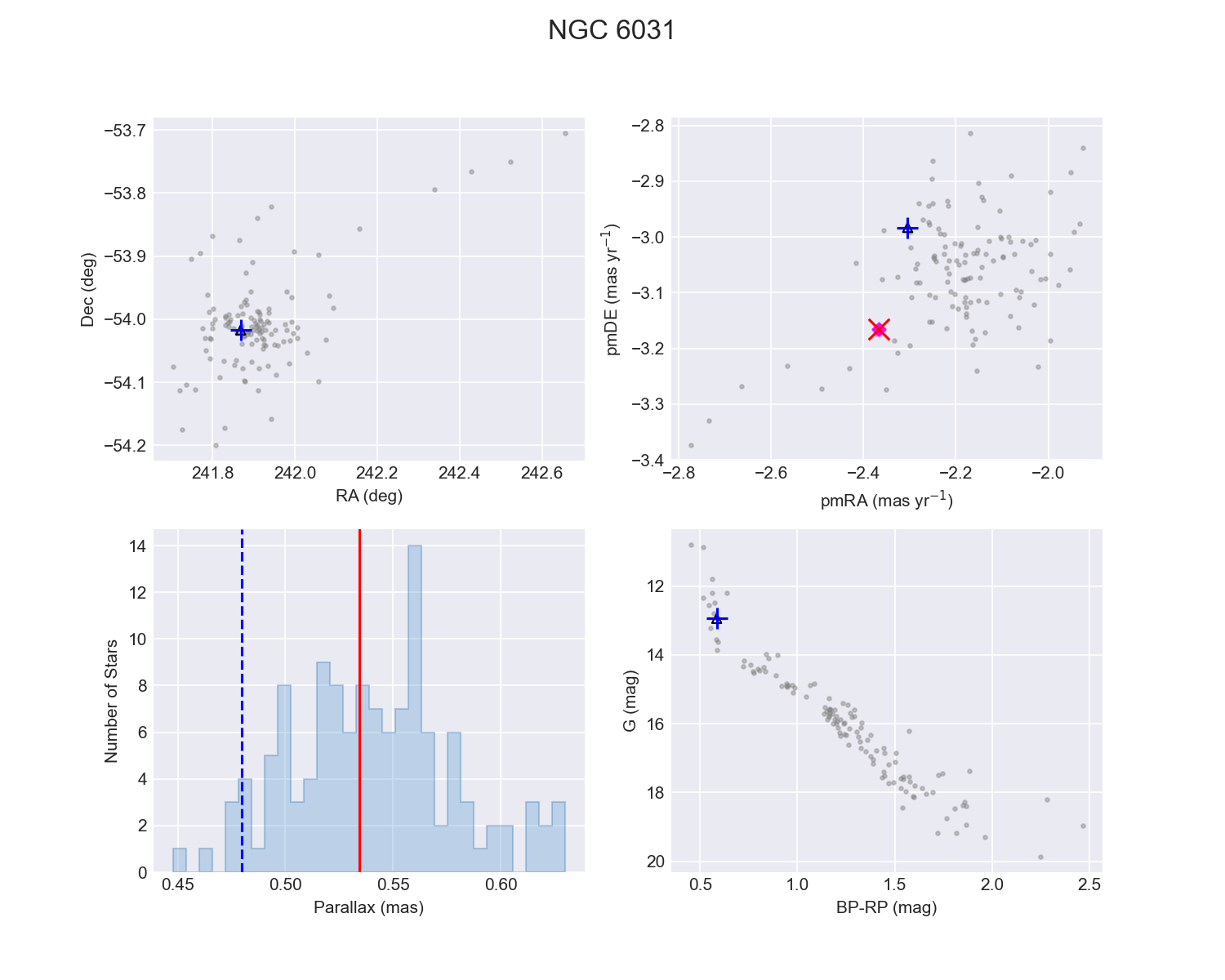}
    \includegraphics[width=0.45\linewidth]{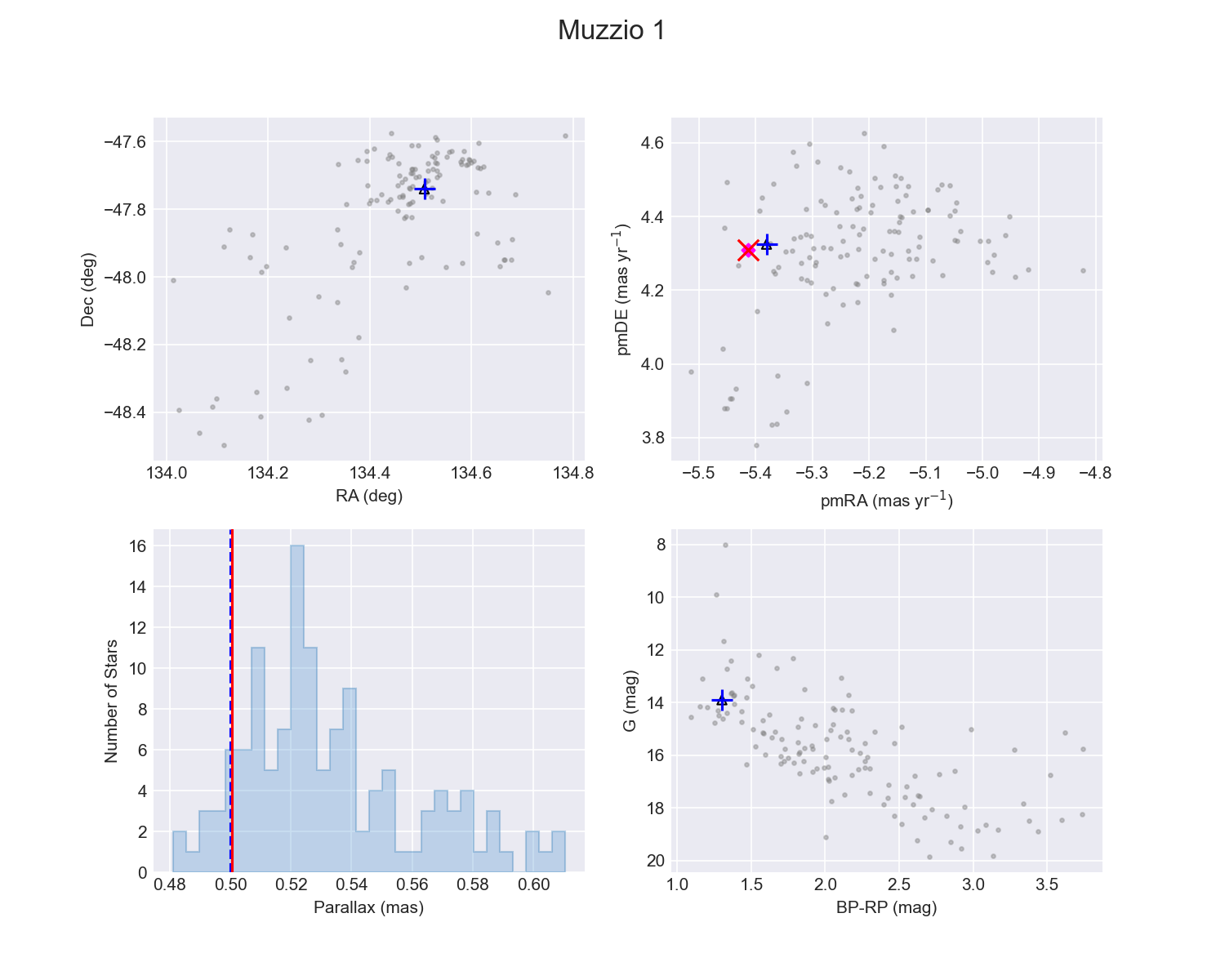}
    \caption{Distributions of the three Class~III sources within their host clusters. Each subpanel shows celestial positions (top left), a proper-motion diagram (top right), a parallax distribution (bottom left), and a color–color diagram (bottom right). The target sources are marked by blue plus sign, other astrometric binaries in the same cluster by open black triangles, and other cluster members by gray points. In the proper-motion panel, the values from the Gaia DR3 \texttt{gaia\_source} table and the NSS orbital solutions are shown as open black triangles and magenta diamonds, respectively. For the target itself, the corresponding values are highlighted by the blue plus sign and a red cross. In the parallax panel, the red solid and blue dashed vertical lines indicate the target parallax from the \texttt{gaia\_source} table and the NSS orbital solution, respectively.}
    \label{fig:clusters}
\end{figure}

Figure~\ref{fig:clusters} presents the host-cluster distributions for the Class~III candidates listed in Table~\ref{tab:amrf}.  For each cluster, the sky projection, proper-motion distribution, parallax distribution, and color--color diagram are shown. For comparison, we also overplot other astrometric binaries in the same cluster, if available. 

\subsection{SED fitting}
\label{subsect:sed}

To assess the reliability of the visible star masses derived from \citet{Hunt+2024}, multi-band SED fitting was performed on a subset of the sources using the software \texttt{ARIADNE} \citep{Vines+2022}. The photometric data include \gaia\ \citep{GaiaDR3+2023}, Pan-STARRS1 \citep{Chambers+2016}, SDSS \citep{Alam+2015}, TESS \citep{Stassun+2019}, 2MASS ($J$, $H$, $K_s$; \citealt{Skrutskie+2006}), and WISE ($W1$, $W2$; \citealt{Wright+2010}). Priors on metallicity, effective temperature, and surface gravity were taken from LAMOST DR13 \citep{Cui+2012, Luo+2015}. These parameters were used as inputs to the MIST \citep{Dotter+2016} isochrone models, yielding reliable SED mass estimates for 78 sources. 

\begin{figure}[htbp]
    \centering
    \begin{subfigure}[]{0.45\linewidth}
        \centering
        \includegraphics[width=\linewidth]{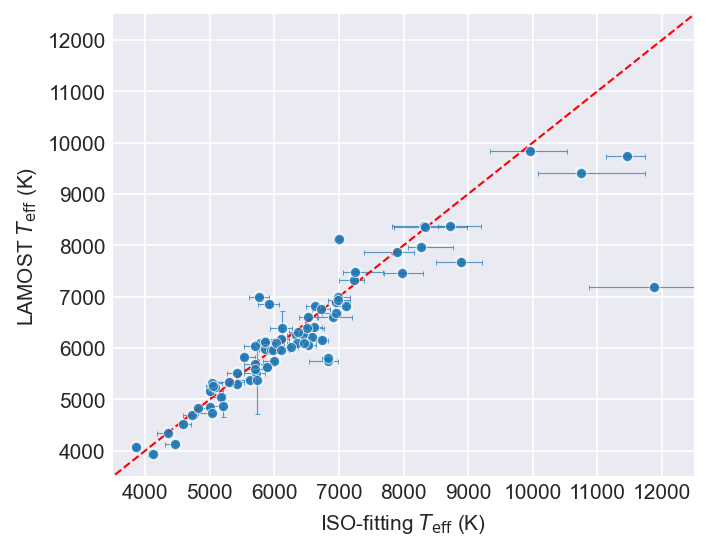}
        \caption{}
        \label{fig:teff}
    \end{subfigure}
    \begin{subfigure}[]{0.45\linewidth}
        \centering
        \includegraphics[width=\linewidth]{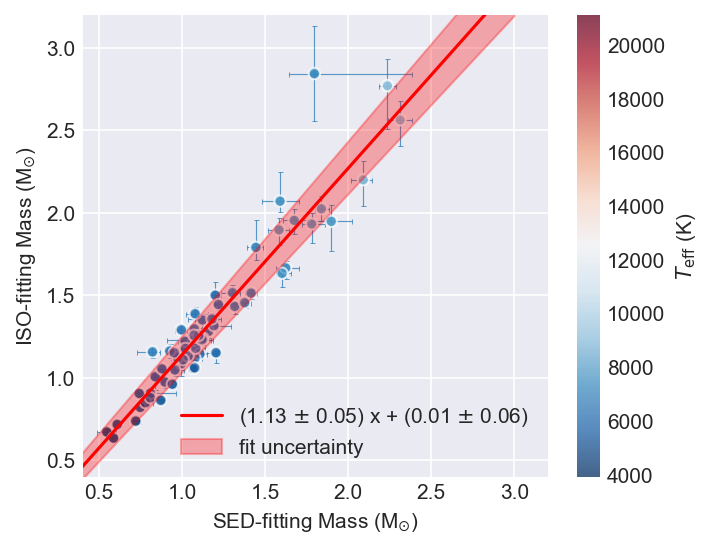}
        \caption{}
        \label{fig:iso_vs_sed}
    \end{subfigure}
    \caption{\textit{Left:} Comparison of effective temperature derived from ISO-cfitting and those from LAMOST. \textit{Right:} Comparison of masses derived from cluster isochrone fitting by \citet{Hunt+2024} versus those from SED fitting. The best-fit linear relation is $(1.13 \pm 0.05)\,x + (0.01 \pm 0.06)$, shown as the red line with the red shaded region indicating the fit uncertainty.}
    \label{fig:placeholder}
\end{figure}

Figure~\ref{fig:teff} compares the ISO-derived effective temperatures with those from LAMOST DR13. The effective temperatures derived from these two independent methods exhibit good consistency. 
Figure~\ref{fig:iso_vs_sed} compares the SED-derived masses with the isochrone masses from \citet{Hunt+2024} catalog. A linear fit yields $(1.13 \pm 0.05)x+(0.01 \pm 0.06)$, with an RMS of 0.15, indicating good consistency.


\subsection{Joint \gaia\ Astrometric and RV Orbit Fitting}
\label{subsect:mcmc_fit}
The joint \gaia\ astrometric and RV orbit fitting procedure described in \citet{El-Badry+2023a,El-Badry+2023b} was followed for sources with multi-epoch RVs. MCMC fitting was conducted using \texttt{emcee} \citep{Foreman+2013}. Upon completion of the fitting, the parametric orbit was constructed from the marginal posterior median parameters. The quality metrics include the reduced chi-squared ($\chi^2$), the fraction of epochs with residuals small than $2\sigma$ ($f_{2\sigma}$), and the median of the absolute normalized residuals ($z$). The chi-squared is defined as \begin{equation}\label{eq:chi2}
  \chi^2 = \frac{1}{N} \sum_{i=1}^{N}
  \left(\frac{v_{i,\mathrm{obs}} - v_{i,\mathrm{mod}}}{\sigma_{v,i}}
  \right)^{\!2},
\end{equation}
where $N$ is the total number of observational epochs, $v_{i,\mathrm{obs}}$ and $v_{i,\mathrm{mod}}$ are the observed and modeled RVs for epoch~$i$, and $\sigma_{v,i}$ is the observational uncertainty for epoch~$i$. Since the joint \gaia+RV fit involves 9~orbital parameters that are not independently constrained by the $N$~RV epochs alone, we define $\chi^2$ with a prefactor of $1/N$ rather than $1/(N-9)$.

The metrics $f_{2\sigma}$ and $z$ are defined as
\begin{equation}
    f_{2\sigma} = \frac{1}{N}\sum_{i=1}^{N}\left(\left| \frac{v_{i, \rm obs} - v_{i, \rm mod}}{\sigma_{v, i}}  \right| \le2 \right),
\end{equation}
\begin{equation}\label{eq:z}
  z = \mathrm{median}\left|\frac{v_{i,\mathrm{obs}} -
  v_{i,\mathrm{mod}}}{\sigma_{v,i}}\right|.
\end{equation}

\begin{figure}[htbp]
    \centering
    \includegraphics[width=0.9\linewidth]{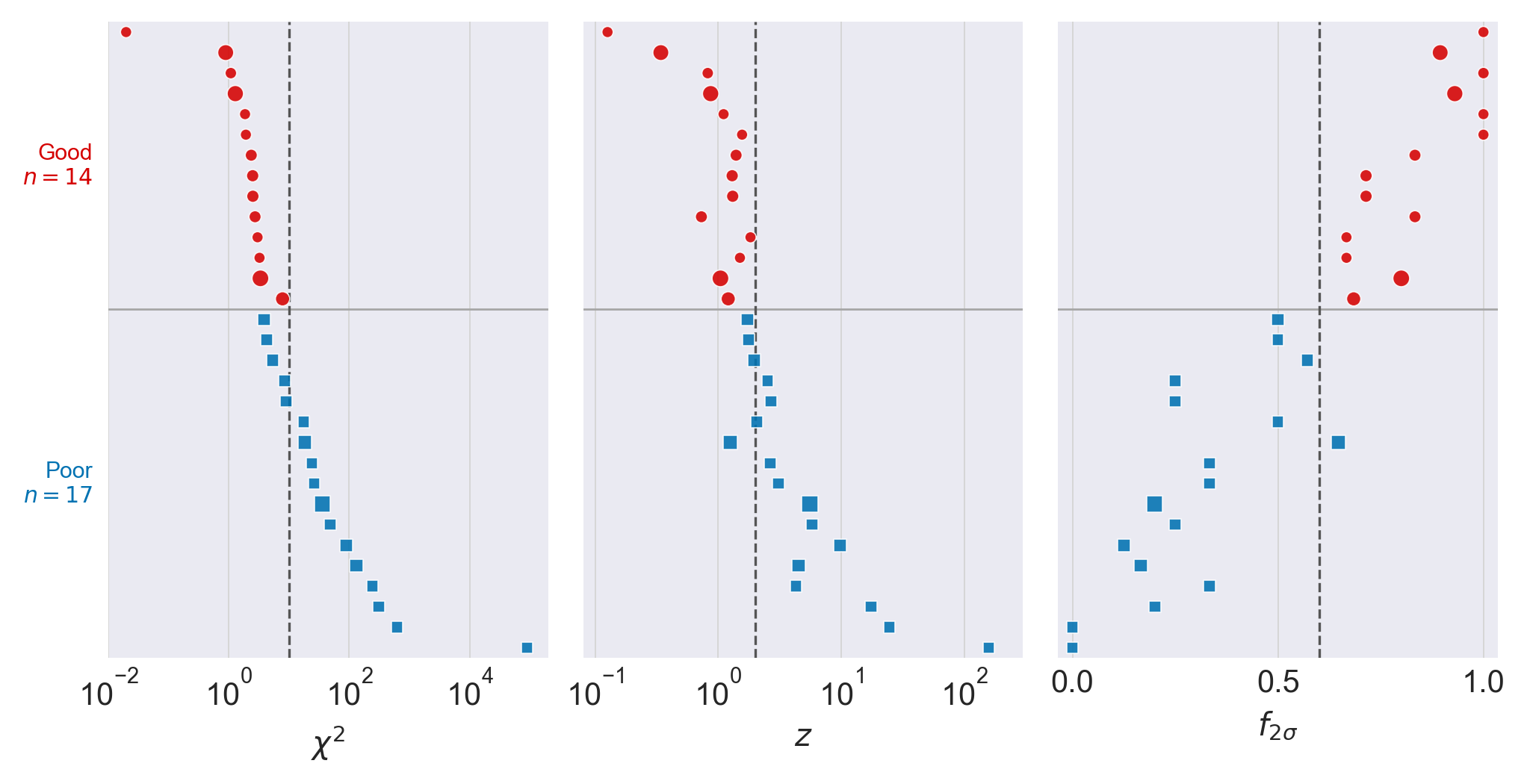}
    \caption{Fit quality for the 31 multi-epoch RV sources.  From left to right: distributions of $\chi^2$, $z$, and $f_{2\sigma}$. Each row corresponds to a single source.  Colors and symbol shapes indicate the quality grade: red circles = good, blue squares = poor.  Point size scale with the number of epochs per source. The dashed vertical lines in each panel mark adopted thresholds: $\chi^2 = 10$ (left), $z = 2$ (middle), $f_{2\sigma} = 0.6$ (right). }
    \label{fig:mcmc_summary}
\end{figure}

\begin{figure}
    \centering
    \includegraphics[width=0.45\linewidth]{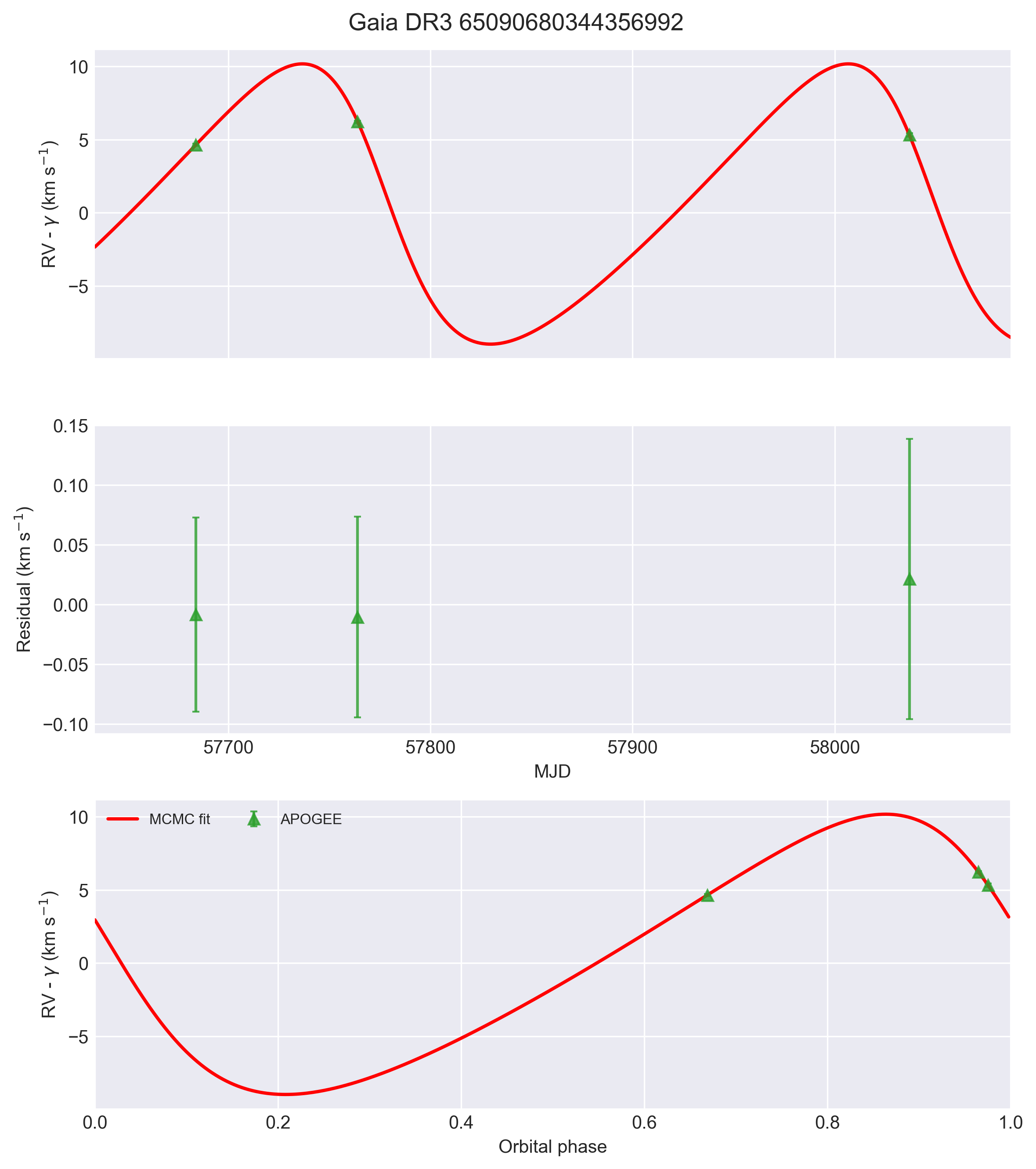}
    \includegraphics[width=0.45\linewidth]{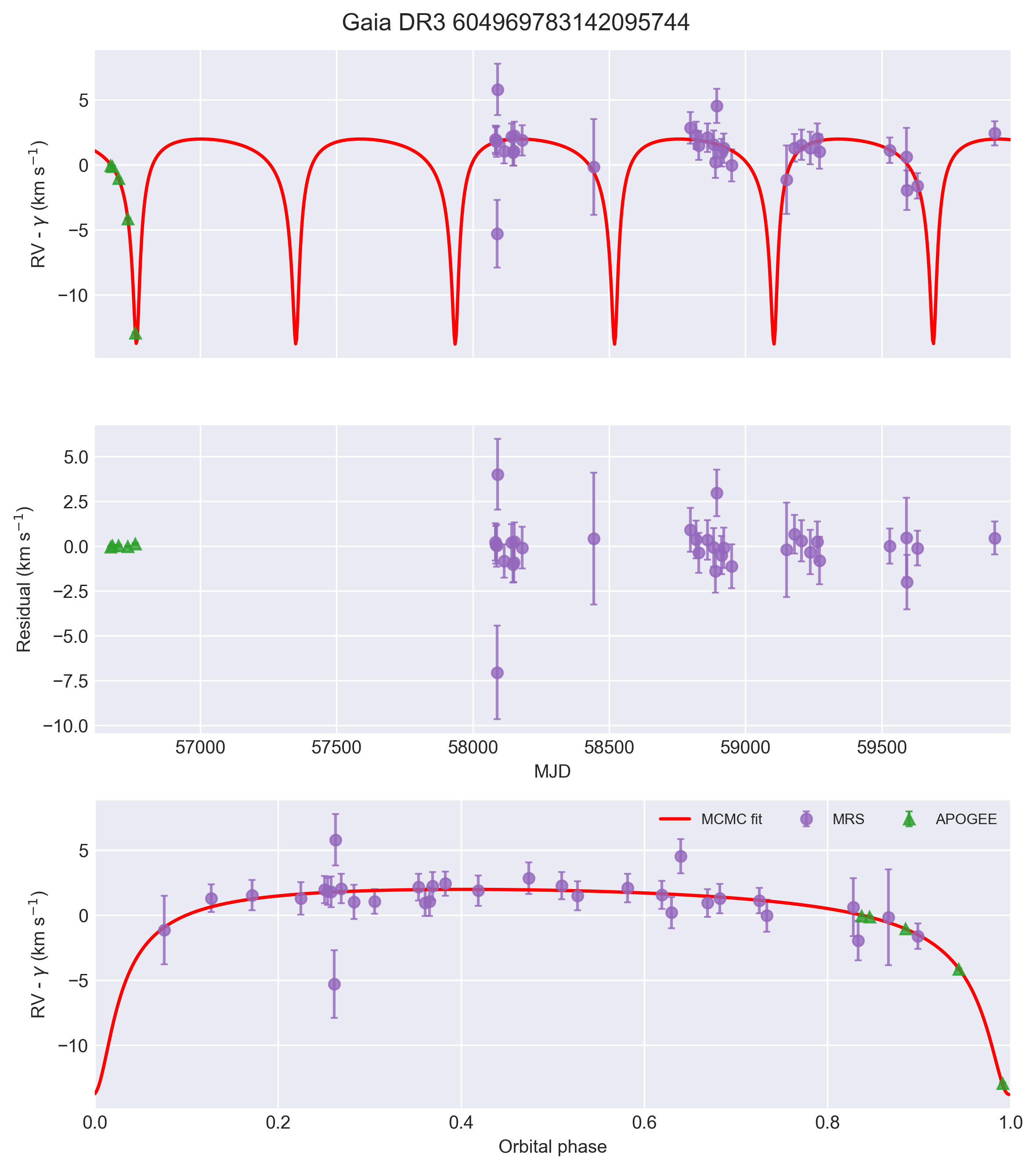}
    \includegraphics[width=0.45\linewidth]{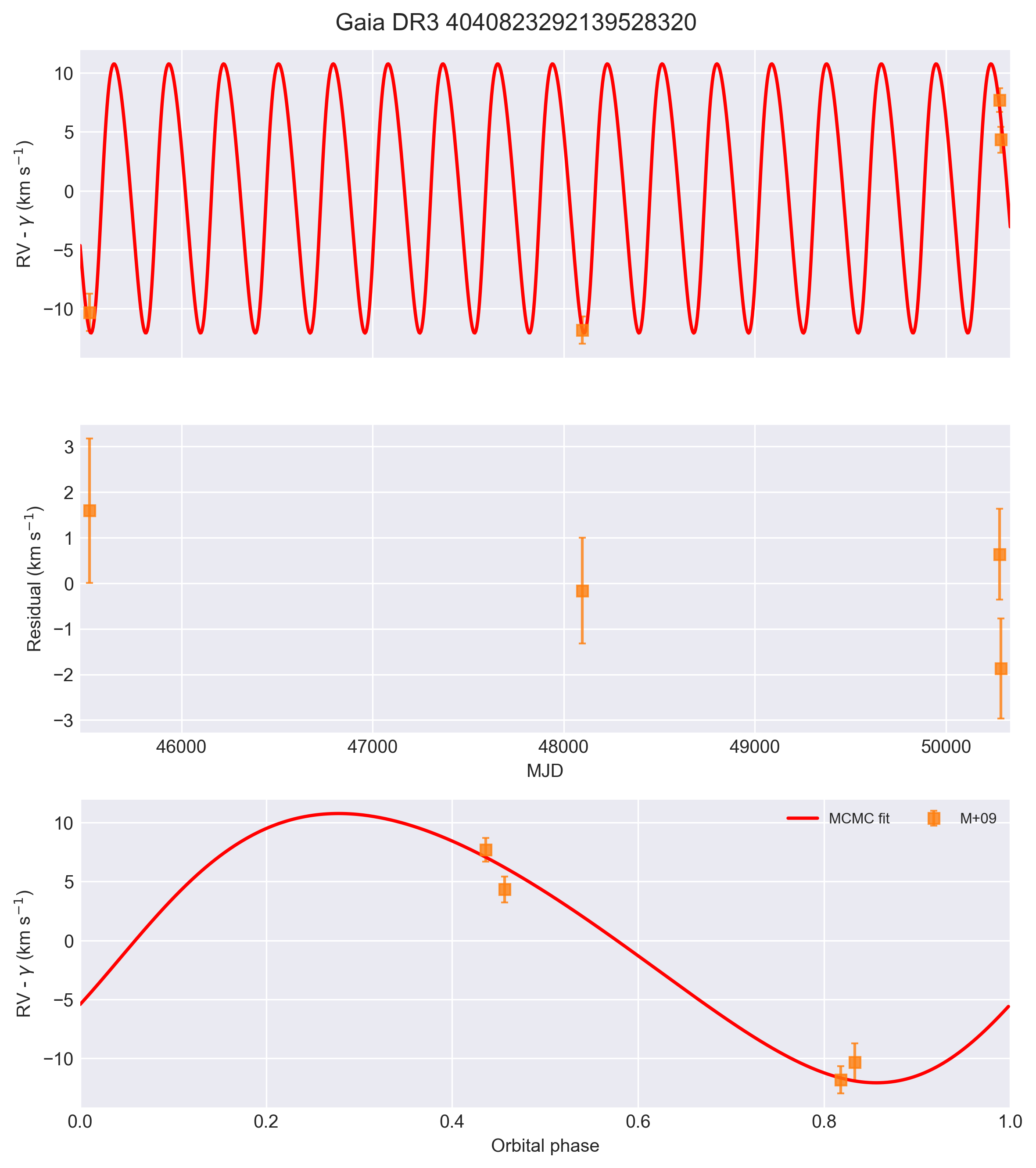}
    \includegraphics[width=0.45\linewidth]{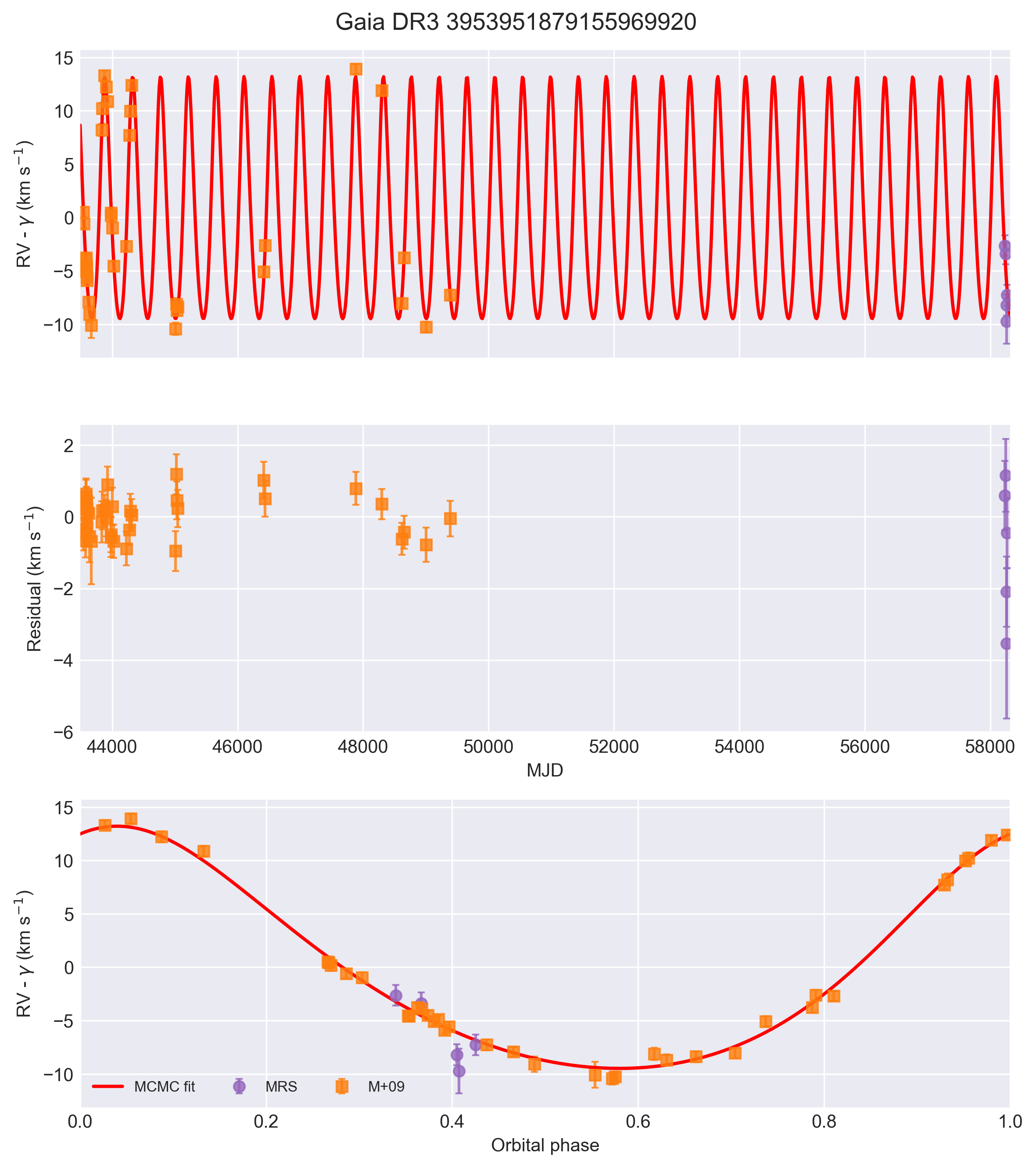}
    \caption{Multi-epoch RV measurements and best-fit \gaia\ + RV joint solutions. Each panel (from top to bottom) displays the RV data overplotted with the fitted model (red solid curve), the corresponding residuals, and the phase-folded RV curve. Green triangles denote data from APOGEE, orange squares denote data from \citet{Mermilliod+2009}, purple points denote data from LAMOST medium-resolution, respectively. }
    \label{fig:mcmc_fit_1}
\end{figure}

\begin{figure}
    \centering
    \includegraphics[width=0.45\linewidth]{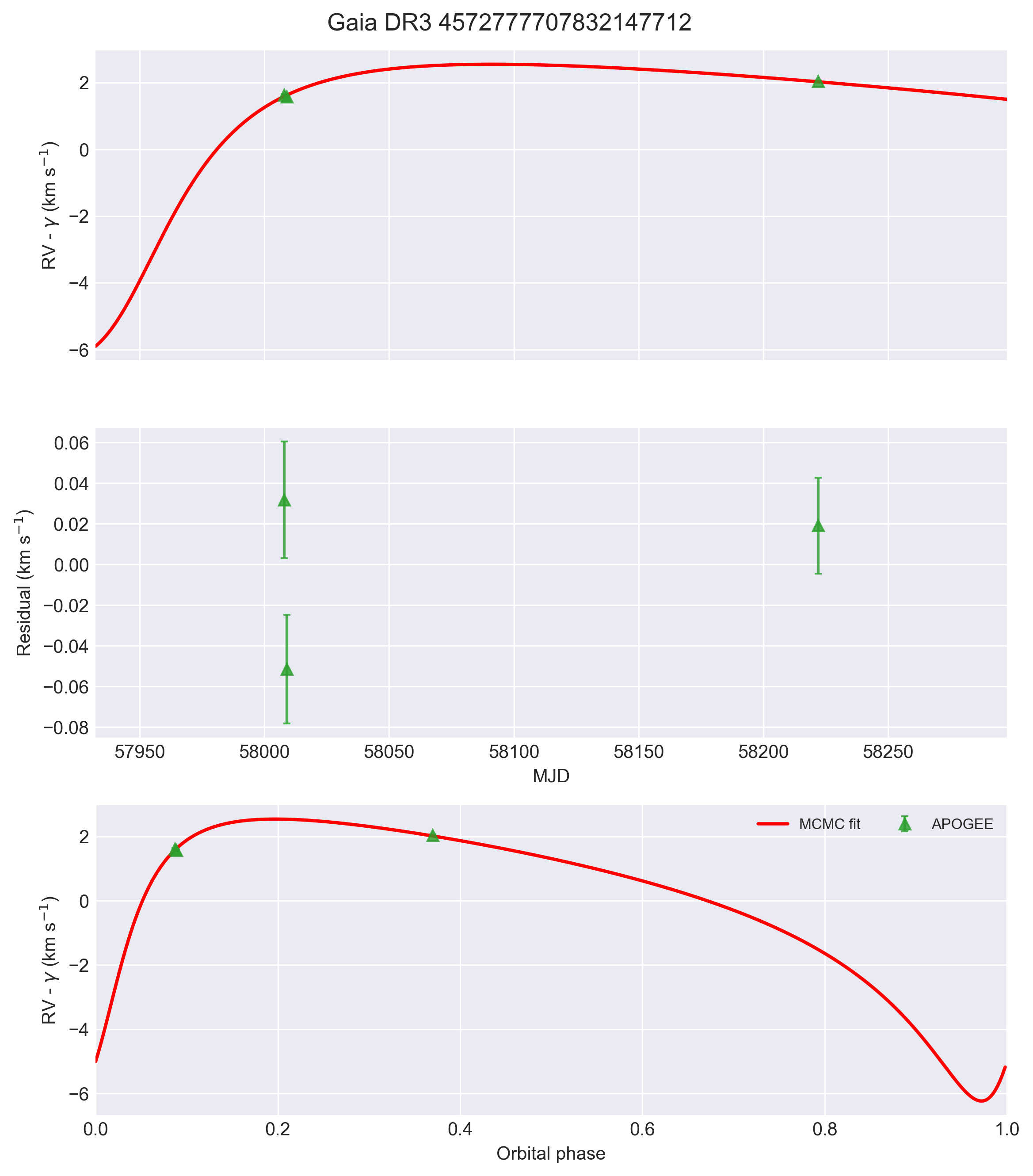}
    \includegraphics[width=0.45\linewidth]{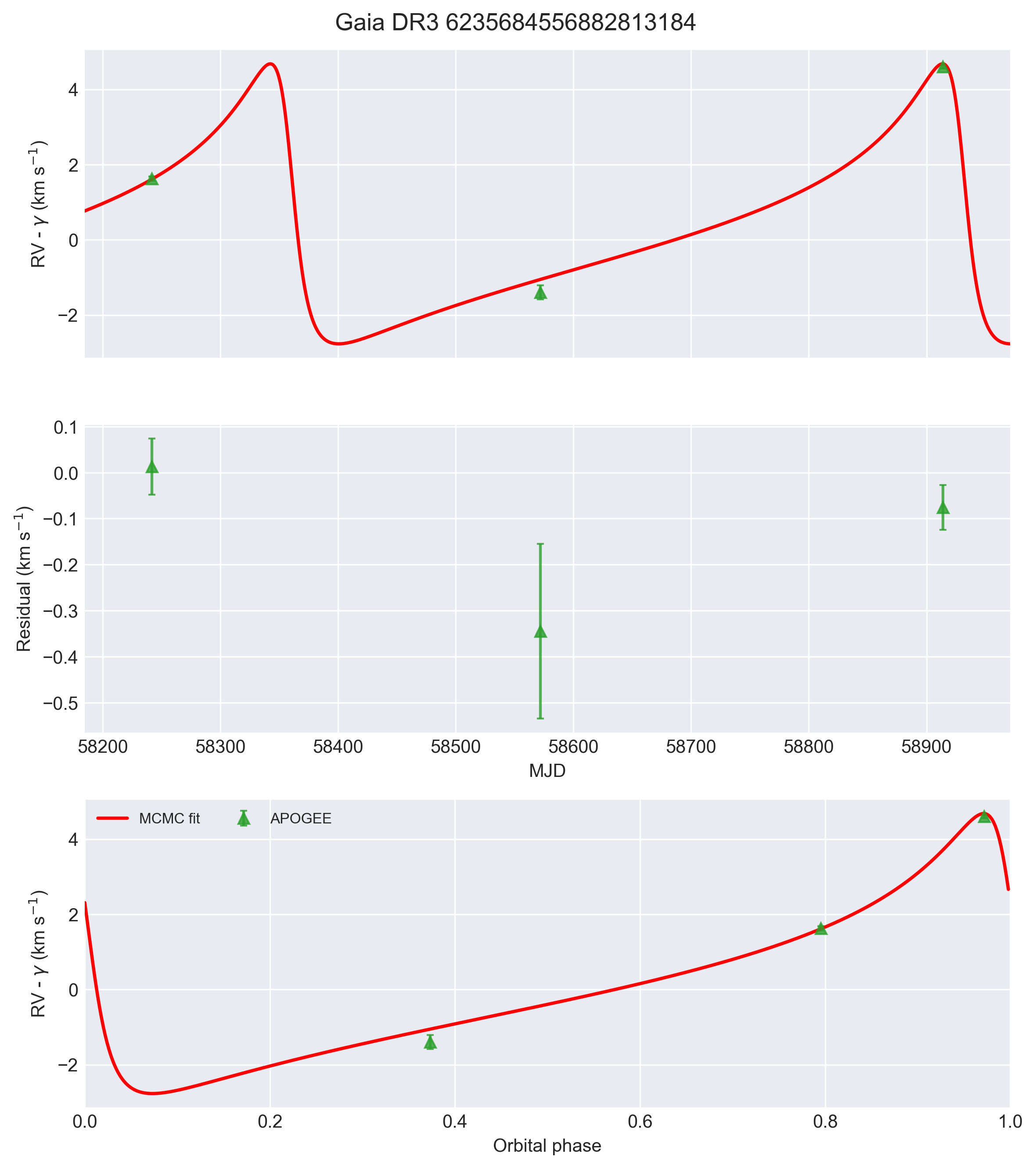}
    \includegraphics[width=0.45\linewidth]{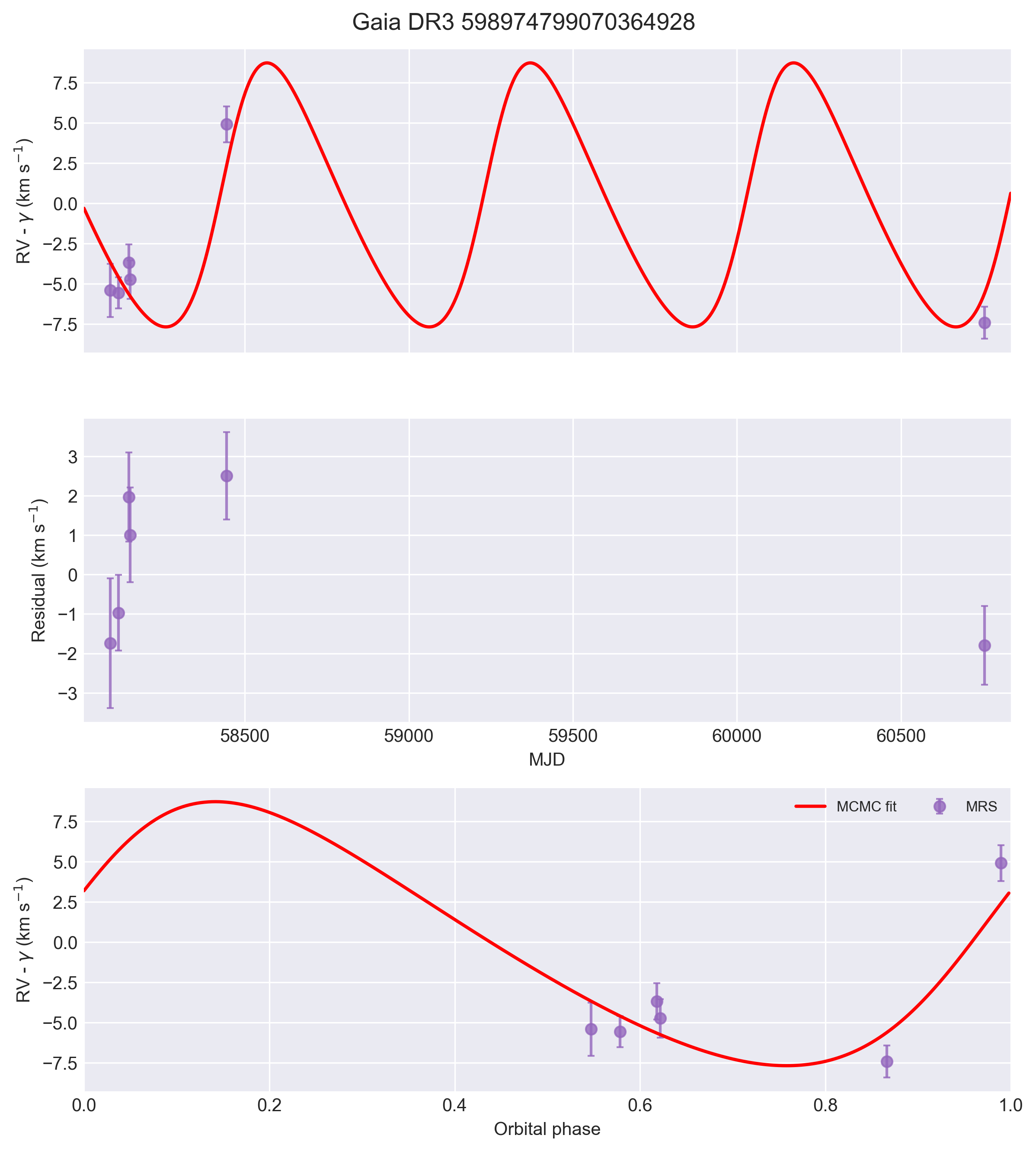}
    \includegraphics[width=0.45\linewidth]{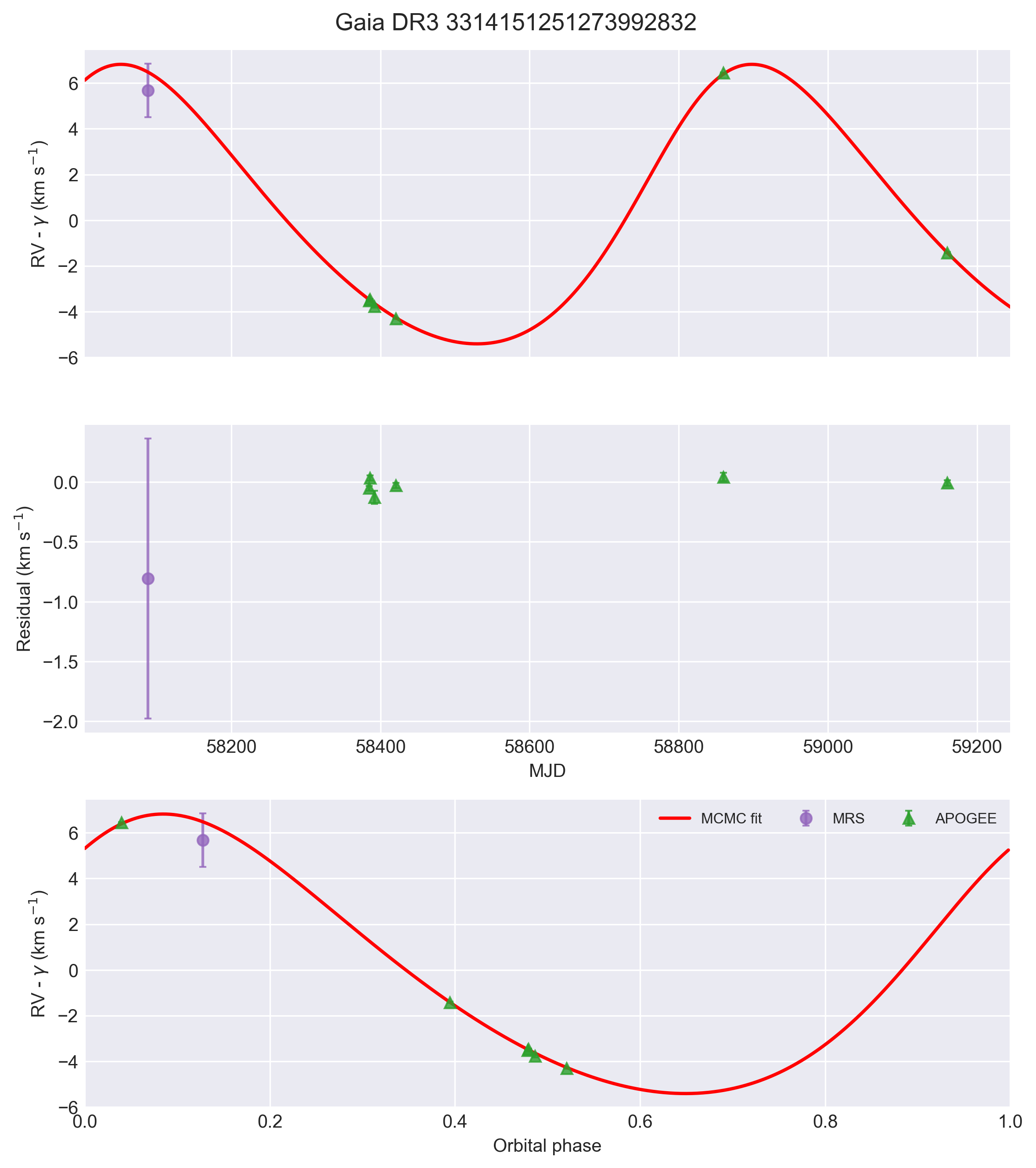}
    \caption{Continue}
    \label{fig:mcmc_fit_3}
\end{figure}

\begin{figure}
    \centering
    \includegraphics[width=0.45\linewidth]{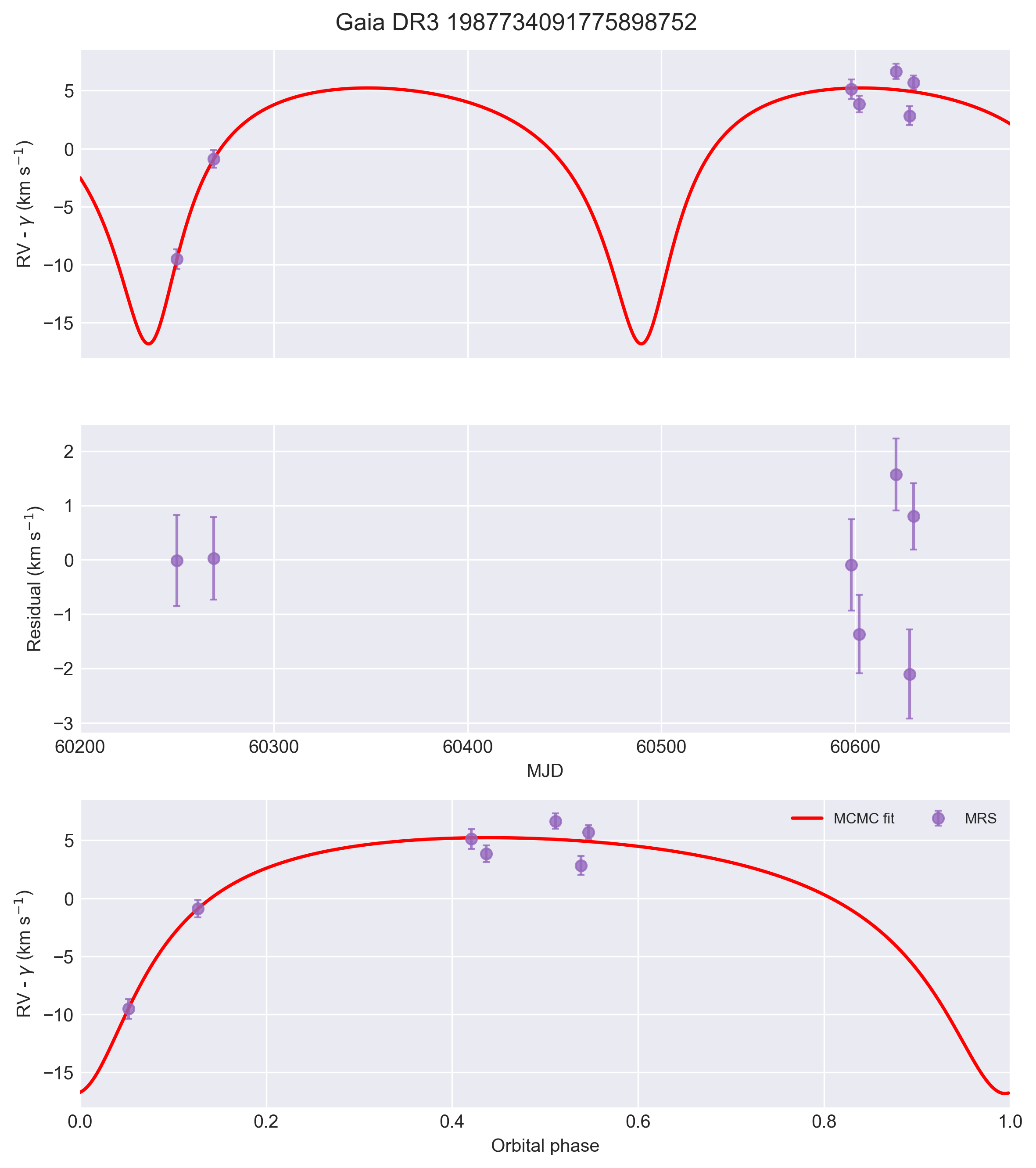}
    \includegraphics[width=0.45\linewidth]{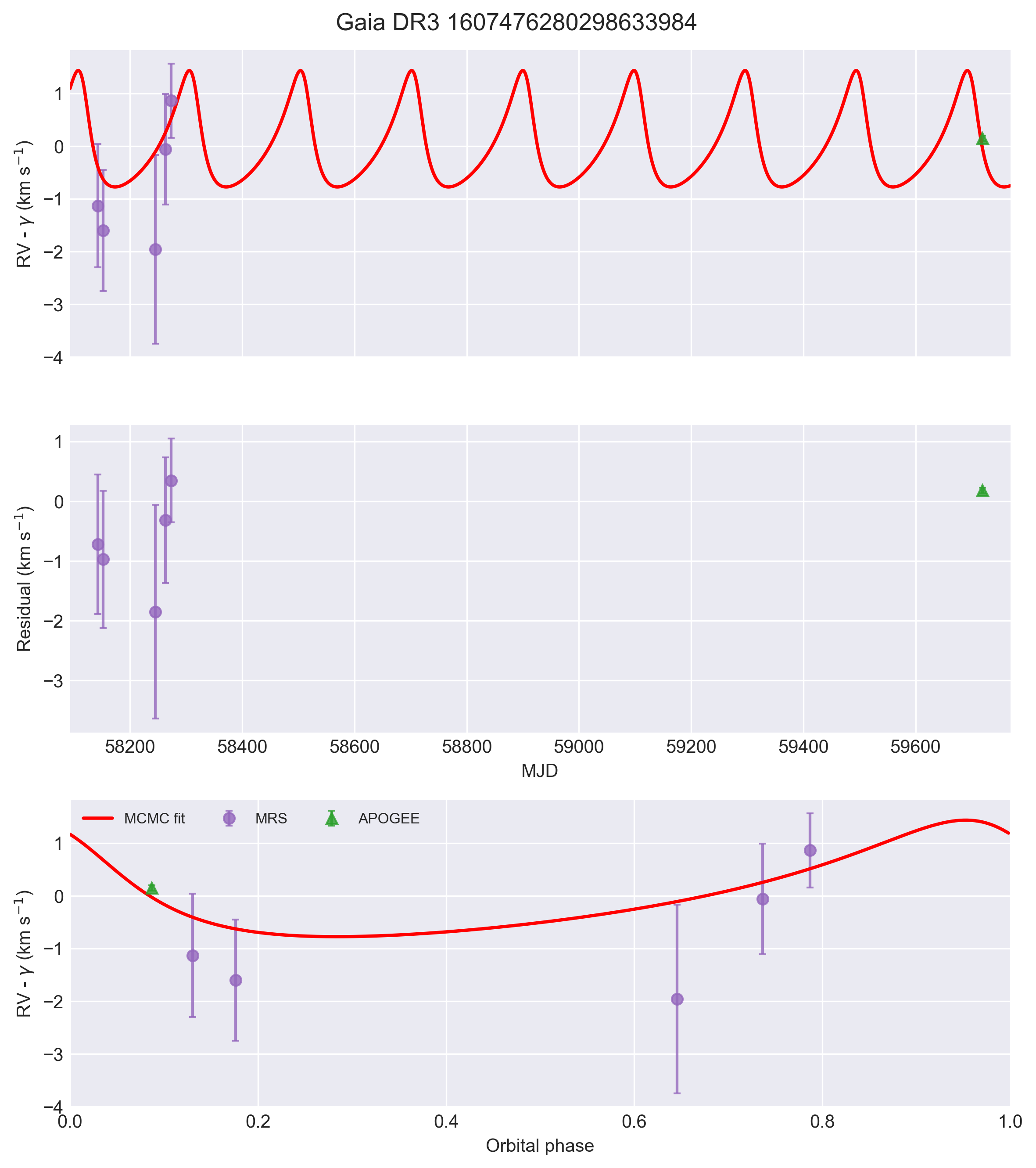}
    \includegraphics[width=0.45\linewidth]{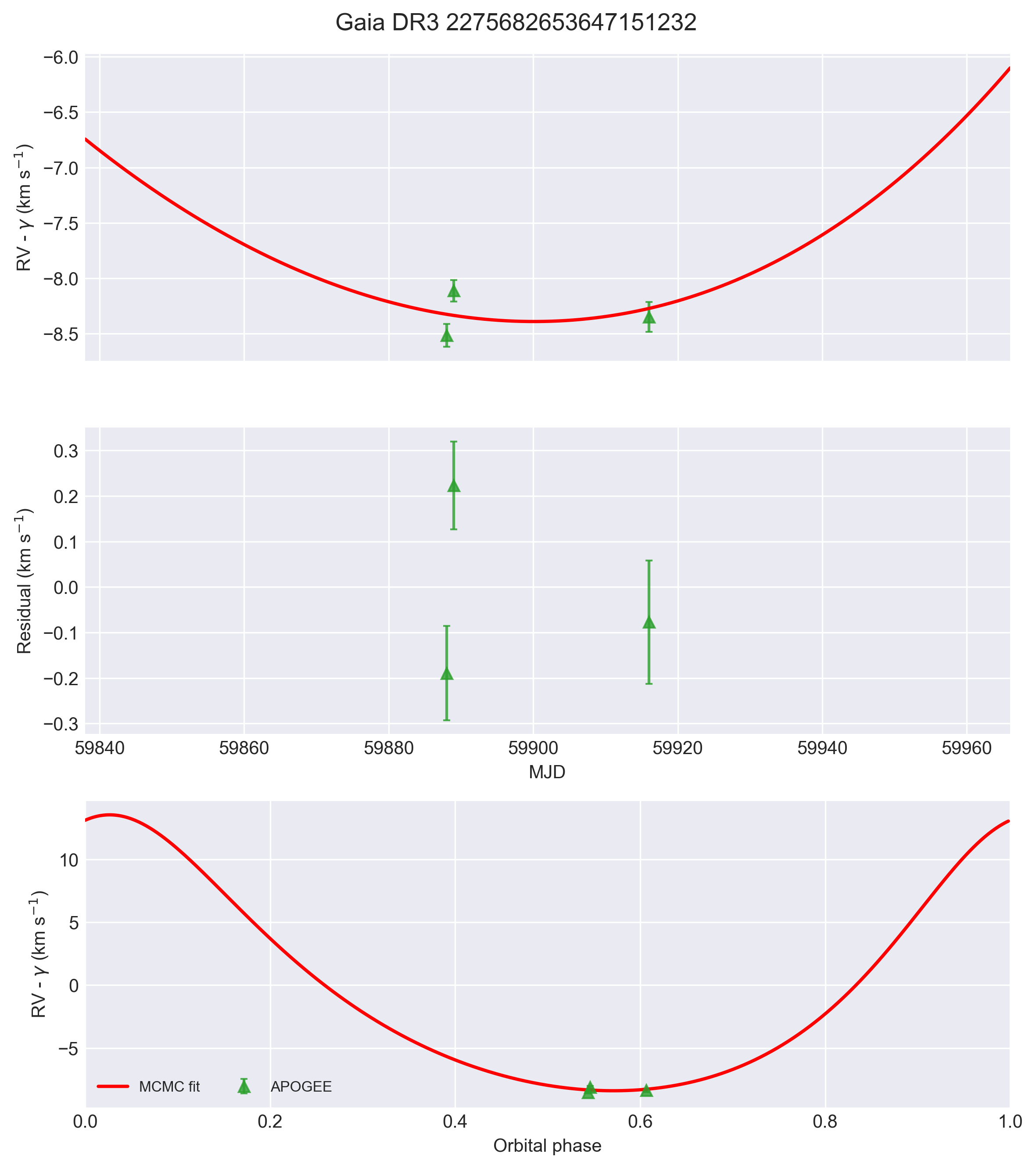}
    \includegraphics[width=0.45\linewidth]{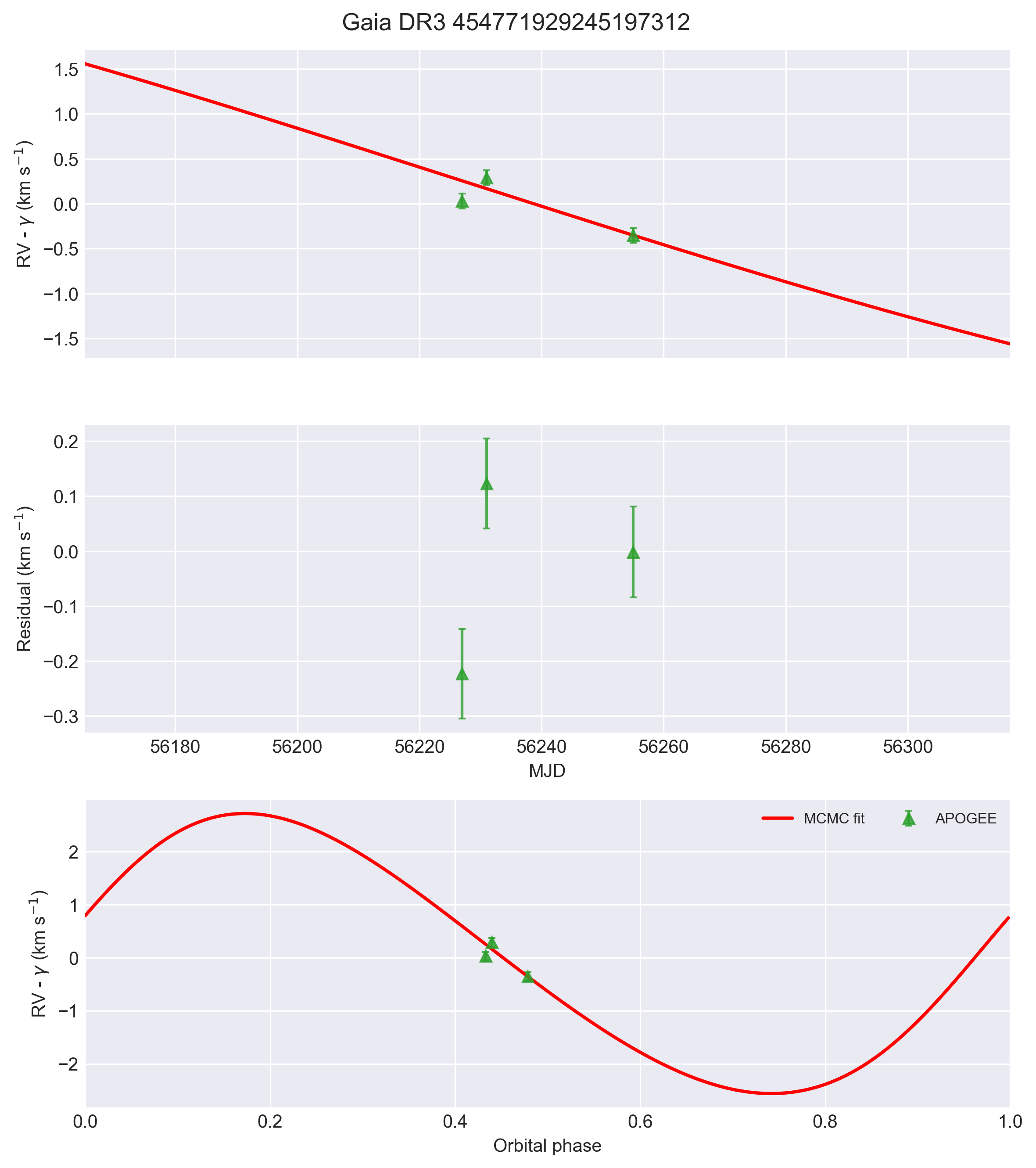}
    \caption{Continue}
    \label{fig:mcmc_fit_4}
\end{figure}

\begin{figure}
    \centering
    \includegraphics[width=0.45\linewidth]{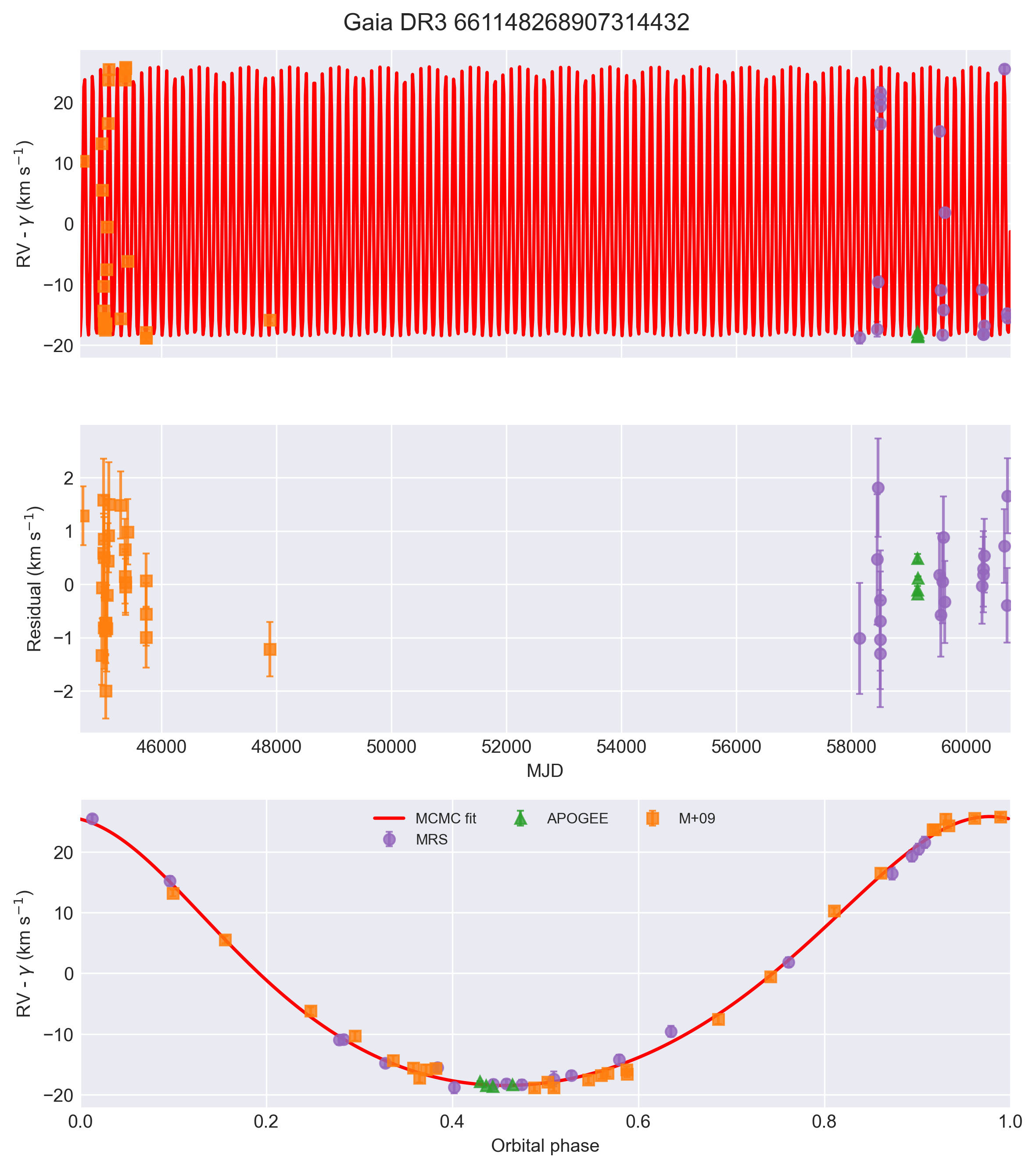}
    \includegraphics[width=0.45\linewidth]{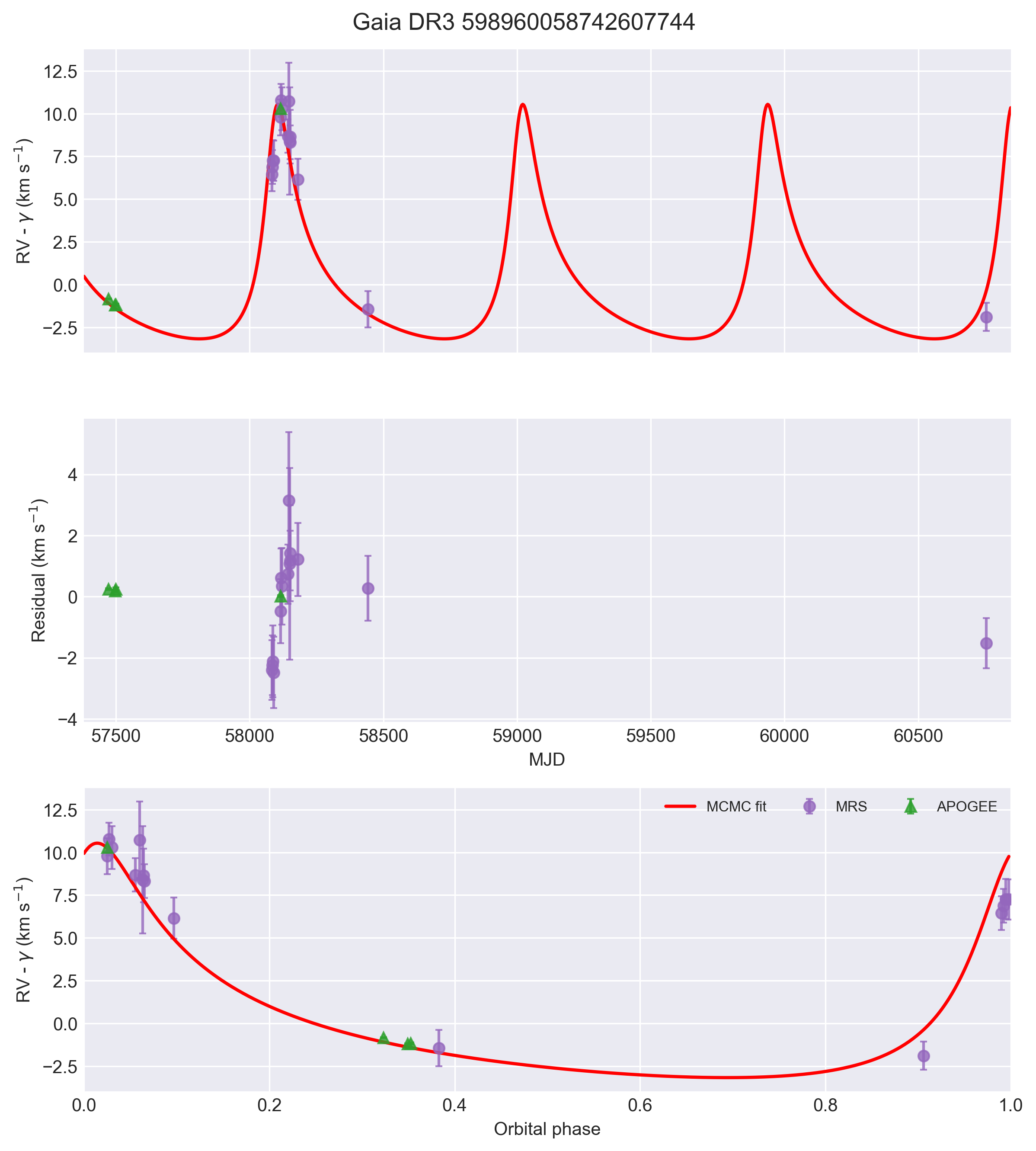}
    \caption{Continue}
    \label{fig:mcmc_fit_5}
\end{figure}

The quality classification criteria used in this work were $\chi^2 \leq 10$, $z \leq 2$, and $f_{2\sigma} \geq 0.6$. Figure~\ref{fig:mcmc_summary} shows the distribution of these criteria across the 31 multi-epoch RV sources. Table~\ref{tab:mcmc_good} lists the systems that meet good results from the joint \gaia+RV orbit fitting. Figures~\ref{fig:mcmc_fit_1} to \ref{fig:mcmc_fit_5} present the milti-epoch RV measurements and the best-fit \gaia+RV joint solution for sources with a good joint-fit quality.

\begin{table}[!ht]
    \tiny
    \centering
    \caption{Good Results from Joint \gaia+RV Orbit
    Fitting \label{tab:mcmc_good}}
    \setlength{\tabcolsep}{3pt} 
    \begin{tabular}{cccccccccccc}
        \hline \hline
         Source ID & NSS type & Cluster & Epochs & $P$ &$i$ & $\Omega$ & $\omega$ & $T_{p}$ & $\gamma$ & $M_{1}^{*}$ & $M_{2}^{*}$   \\ 
         & & & & (day) & & & & & & ($\msol$) & ($\msol$) \\ \hline
        65090680344356992 & Orbital & Melotte 22 & 3 & 270.1$^{+0.5}_{-0.5}$ & 1.92$^{+0.07}_{-0.07}$ & 1.21$^{+0.07}_{-0.06}$ & 1.32$^{+0.12}_{-0.11}$ & 57503.36$^{+4.60}_{-4.18}$ & 5.47$^{+0.26}_{-0.26}$ & 1.36$^{+0.29}_{-0.29}$ & 0.44$^{+0.06}_{-0.06}$ \\ 
        604969783142095744 & Orbital & NGC 2682 & 38 & 584.7$^{+3.1}_{-2.2}$ & 1.95$^{+0.08}_{-0.10}$ & 1.21$^{+0.06}_{-0.06}$ & -3.01$^{+0.05}_{-0.05}$ & 57351.64$^{+3.07}_{-3.49}$ & 34.13$^{+0.24}_{-0.24}$ & 1.45$^{+0.32}_{-0.30}$ & 0.32$^{+0.05}_{-0.05}$ \\ 
        4040823292139528320 & Orbital & NGC 6475 & 4 & 286.8$^{+0.8}_{-0.9}$ & 1.13$^{+0.16}_{-0.14}$ & 2.95$^{+0.26}_{-0.29}$ & -2.00$^{+0.41}_{-0.29}$ & 57326.26$^{+14.95}_{-14.66}$ & -17.23$^{+1.36}_{-1.93}$ & 1.47$^{+0.30}_{-0.30}$ & 0.64$^{+0.10}_{-0.10}$ \\ 
        3953951879155969920 & AstroSpectroSB1 & Melotte 111 & 43 & 444.3$^{+0.1}_{-0.1}$ & 1.17$^{+0.03}_{-0.03}$ & -1.95$^{+0.03}_{-0.03}$ & -0.36$^{+0.03}_{-0.03}$ & 57193.50$^{+3.33}_{-3.25}$ & -0.38$^{+0.08}_{-0.08}$ & 1.30$^{+0.29}_{-0.29}$ & 0.69$^{+0.09}_{-0.09}$ \\ 
        4572777707832147712 & Orbital & HSC 453 & 3 & 756.8$^{+1.8}_{-1.8}$ & 2.14$^{+0.01}_{-0.01}$ & 0.52$^{+0.01}_{-0.01}$ & -2.37$^{+0.01}_{-0.01}$ & 57185.32$^{+0.98}_{-1.02}$ & -25.79$^{+0.04}_{-0.04}$ & 1.05$^{+0.28}_{-0.29}$ & 0.21$^{+0.03}_{-0.04}$ \\ 
        6235684556882813184 & Orbital & CWNU 1143 & 3 & 571.2$^{+2.0}_{-1.8}$ & 2.24$^{+0.03}_{-0.03}$ & -1.56$^{+0.05}_{-0.06}$ & 1.20$^{+0.07}_{-0.07}$ & 57216.50$^{+2.65}_{-2.70}$ & -3.25$^{+0.13}_{-0.13}$ & 0.69$^{+0.28}_{-0.27}$ & 0.11$^{+0.03}_{-0.03}$ \\ 
        598974799070364928 & Orbital & NGC 2682 & 6 & 802.6$^{+9.5}_{-14.9}$ & 1.76$^{+0.40}_{-0.25}$ & 0.86$^{+0.26}_{-0.13}$ & -1.24$^{+0.28}_{-0.25}$ & 56848.22$^{+51.95}_{-32.13}$ & 36.91$^{+0.76}_{-0.75}$ & 1.35$^{+0.29}_{-0.29}$ & 0.55$^{+0.10}_{-0.09}$ \\ 
        3314151251273992832 & Orbital & Melotte 25 & 7 & 846.1$^{+1.1}_{-1.0}$ & 0.94$^{+0.01}_{-0.01}$ & -2.43$^{+0.01}_{-0.01}$ & -0.72$^{+0.01}_{-0.01}$ & 57134.03$^{+2.55}_{-2.58}$ & 40.16$^{+0.03}_{-0.03}$ & 0.96$^{+0.33}_{-0.29}$ & 0.41$^{+0.08}_{-0.08}$ \\ 
        1987734091775898752 & Orbital & UPK 167 & 7 & 254.3$^{+0.4}_{-0.4}$ & 2.29$^{+0.07}_{-0.06}$ & -0.05$^{+0.10}_{-0.11}$ & -3.00$^{+0.12}_{-0.12}$ & 57439.93$^{+4.05}_{-3.97}$ & -14.12$^{+0.56}_{-0.56}$ & 1.49$^{+0.30}_{-0.30}$ & 0.61$^{+0.08}_{-0.08}$ \\ 
        1607476280298633984 & Orbital & HSC 759 & 6 & 198.1$^{+1.2}_{-1.6}$ & 0.60$^{+0.53}_{-0.43}$ & -0.83$^{+0.40}_{-1.12}$ & 0.71$^{+1.03}_{-0.44}$ & 57324.83$^{+7.68}_{-7.87}$ & -5.87$^{+0.34}_{-0.53}$ & 1.13$^{+0.30}_{-0.29}$ & 0.06$^{+0.02}_{-0.01}$ \\ 
        2275682653647151232 & AstroSpectroSB1 & HSC 976 & 3 & 444.0$^{+0.7}_{-0.6}$ & 1.14$^{+0.02}_{-0.02}$ & -1.63$^{+0.04}_{-0.04}$ & -0.28$^{+0.05}_{-0.05}$ & 57426.86$^{+3.93}_{-3.75}$ & 4.55$^{+0.29}_{-0.28}$ & 1.43$^{+0.29}_{-0.30}$ & 0.69$^{+0.08}_{-0.09}$ \\ 
        454771929245197312 & Orbital & CWNU 1076 & 3 & 617.5$^{+4.9}_{-5.4}$ & 0.51$^{+0.18}_{-0.14}$ & -0.45$^{+0.94}_{-0.77}$ & -1.30$^{+0.72}_{-0.93}$ & 57194.57$^{+37.25}_{-37.73}$ & 2.08$^{+1.50}_{-1.94}$ & 1.91$^{+0.30}_{-0.34}$ & 0.38$^{+0.05}_{-0.05}$ \\ 
        661148268907314432 & AstroSpectroSB1 & NGC 2632 & 50 & 142.9$^{+0.0}_{-0.0}$ & 1.65$^{+0.02}_{-0.02}$ & 2.94$^{+0.02}_{-0.02}$ & 0.20$^{+0.03}_{-0.03}$ & 57378.22$^{+0.69}_{-0.70}$ & 34.61$^{+0.10}_{-0.11}$ & 1.27$^{+0.30}_{-0.29}$ & 0.90$^{+0.11}_{-0.12}$ \\ 
        598960058742607744 & Orbital & NGC 2682 & 19 & 916.0$^{+41.4}_{-80.6}$ & 0.99$^{+0.09}_{-0.09}$ & -0.05$^{+0.15}_{-0.21}$ & -0.42$^{+0.32}_{-0.18}$ & 57176.09$^{+98.56}_{-38.00}$ & 34.85$^{+0.37}_{-0.35}$ & 1.55$^{+0.34}_{-0.32}$ & 0.48$^{+0.08}_{-0.07}$ \\  \hline  
    \end{tabular}
\end{table}


\bibliography{sample701}{}
\bibliographystyle{aasjournalv7}



\end{document}